\pdfoutput=1
\documentclass[11pt,twoside,a4paper,cmspaper,final,collab]{cms-tdr}
\def\svnVersion{e1765bf}\def\svnDate{2023/06/02}\def\cmsCernNoTag{CERN-EP-2022-033}\def\cmsCernDate{\today}\def\cmsMessage{Published in the Journal of High Energy Physics as \href{http://dx.doi.org/10.1007/JHEP05(2023)227}{\doi{10.1007/JHEP05(2023)227}.}}
\begin{document}\cmsNoteHeader{EXO-19-014}

\newcommand{\PZggx}{\ensuremath{\PZ/\PGg^{*}}\xspace}

\cmsNoteHeader{EXO-19-014} 
\title{Search for heavy resonances and quantum black holes in \texorpdfstring{$\Pe\PGm$}{emu}, \texorpdfstring{$\Pe\PGt$}{etau}, and \texorpdfstring{$\PGm\PGt$}{mutau} final states in proton-proton collisions at \texorpdfstring{$\sqrt{s}=13\TeV$}{sqrt(s) = 13 TeV}}

\author*[inst1]{CMS Collaboration}

\date{\today}

\abstract{
A search is reported for heavy resonances and quantum black holes decaying into $\Pe\PGm$, $\Pe\PGt$, and $\PGm\PGt$ final states in proton-proton collision data recorded by the CMS experiment at the CERN LHC during 2016--2018 at $\sqrt{s}=13\TeV$, corresponding to an integrated luminosity of 138\fbinv.  The $\Pe\PGm$, $\Pe\PGt$, and $\PGm\PGt$ invariant mass spectra are reconstructed, and no evidence is found for physics beyond the standard model. Upper limits are set at 95\% confidence level on the product of the cross section and branching fraction for lepton flavor violating signals. Three benchmark signals are  studied: resonant $\PGt$ sneutrino production in $R$ parity violating supersymmetric models, heavy \PZpr gauge bosons with lepton flavor violating decays, and nonresonant quantum  black hole production in models with extra spatial dimensions. Resonant $\PGt$ sneutrinos  are excluded for masses up to 4.2\TeV in the $\Pe\PGm$ channel, 3.7\TeV in the $\Pe\PGt$ channel, and 3.6\TeV in the $\PGm\PGt$ channel. A \PZpr boson with lepton flavor violating couplings is excluded  up to a mass of 5.0\TeV in the $\Pe\PGm$ channel, up to 4.3\TeV in the $\Pe\PGt$ channel, and up to 4.1\TeV in the $\PGm\PGt$ channel. Quantum black holes in the benchmark model are excluded up to the threshold mass of 5.6\TeV in the $\Pe\PGm$ channel, 5.2\TeV in the $\Pe\PGt$ channel,  and 5.0\TeV in the $\PGm\PGt$ channel. In addition, model-independent limits are  extracted to allow comparisons with  other models for the same final states and  similar event selection requirements. The results of these searches  provide the most stringent limits  available from collider experiments for heavy particles that undergo lepton flavor violating decays.
}

\hypersetup{
pdfauthor={CMS Collaboration},
pdftitle={Search for heavy resonances and quantum black holes in emu, etau, and mutau final states in proton-proton collisions at sqrt(s) = 13 TeV},
pdfsubject={CMS},
pdfkeywords={CMS, BSM, LFV}}

\maketitle 

\section{Introduction}
\label{sec_intro}
Extensions of the standard model (SM) can accommodate heavy particles that undergo lepton flavor
violating (LFV) decays, motivating thereby searches for deviations from the SM in the $\Pe\PGm$, $\Pe\PGt$, and $\PGm\PGt$ final states.
This paper reports a search for such phenomena in these final states in proton-proton ($\Pp\Pp$) collisions at the CERN LHC, 
and is designed to be as model-independent as possible. Additionally, the results are interpreted in terms of 
the characteristics of the following possible states: a {\PGt} sneutrino ($\PSgn_\PGt$), which can be the
lightest SUSY particle (LSP)~\cite{Farrar:1978xj,deCampos:2008av} in $R$ parity violating (RPV)
supersymmetric (SUSY) models~\cite{Barbier:2004ez}, a heavy \PZpr gauge boson in LFV
models~\cite{Langacker:2008yv}, and quantum black holes (QBHs)~\cite{Calmet:2008dg,Meade:2007sz,Dimopoulos:2001hw,Giddings:2001bu}.
This is the first analysis searching for high-mass lepton flavor violating signals using the full Run 2 data set, recorded at $\sqrt{s}=13\TeV$.

In RPV SUSY models, lepton flavor and lepton number are violated at lowest (Born) level in interactions
between fermions and their superpartners. For resonant $\PSgn_\PGt$ signals,
the trilinear RPV part of the superpotential can be expressed as
\begin{linenomath}
\begin{equation}
W_{\text{RPV}} = \frac{1}{2} \lambda_{ijk} L_i L_j \overline{E}_k
+ \lambda'_{ijk} L_i Q_j \overline{D}_k,
\end{equation}
\end{linenomath}
where $i$, $j$, and $k$ are generation indices, $L$ and $Q$ are
the $\text{SU}(2)_L$ doublet superfields of the leptons and quarks, and $\overline{E}$ and $\overline{D}$ are the respective $\text{SU}(2)_L$
singlet superfields of the charged leptons and down-like quarks.
For simplicity, we assume that all RPV couplings vanish, except for
those that are connected to the
production and decay of the  $\PSgn_\PGt$, and we consider a SUSY mass hierarchy with $\PSgn_\PGt$ as the
LSP. In this model, the $\PSgn_\PGt$ can be produced resonantly in $\Pp\Pp$ collisions 
via the 
$\lambda'$ 
coupling, and can decay either into 
dilepton final states via the 
$\lambda$ 
couplings, or into
quarks via
$\lambda'$ couplings. We consider only the final states with two charged leptons.
Also, this analysis considers $\PSgn_\PGt$ that decay promptly and are not
long-lived~\cite{Graham:2012th}, which we define as having a transverse displacement less than 1\mm from the production vertex.
As long as the $\lambda$ coupling is larger than $10^{-7}$, a $\PSgn_\PGt$ of mass 1\TeV will not be 
considered to be long-lived~\cite{Barbier:2004ez}. 

An extension of the SM through the addition of an extra U(1) gauge symmetry provides a massive \PZpr vector boson~\cite{Langacker:2008yv}.
In our search, we assume that the \PZpr boson has identical couplings to SM particles as the SM \PZ boson, but that the \PZpr boson can also decay to the LFV {\Pe}{\PGm}, {\Pe}{\PGt}, and
{\PGm}{\PGt} final states, each with an assumed branching fraction of 10\%. This value is defined in a benchmark scenario 
commonly used in these searches~\cite{Sirunyan:2018zhy,ATLAS:2018mrn}. 

\begin{figure}[]
\centering
\includegraphics[width=.3\textwidth]{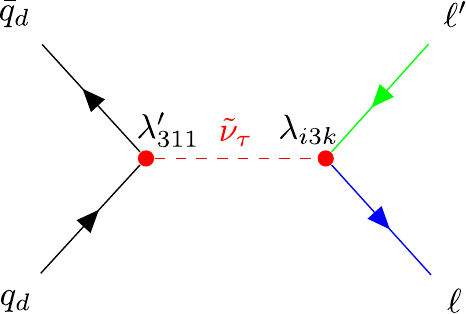}
\includegraphics[width=.3\textwidth]{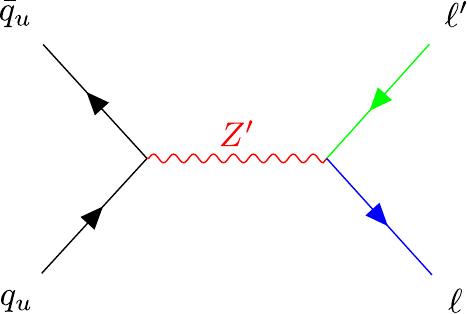}
\includegraphics[width=.3\textwidth]{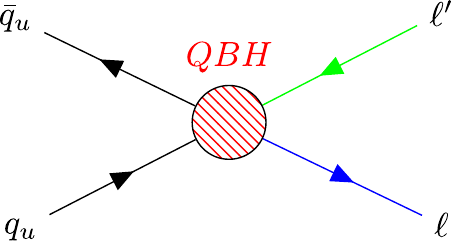}

\caption{Leading order Feynman diagrams considered in our search. 
Left: Resonant production of a $\PGt$ sneutrino in an RPV SUSY model that includes the subsequent decay into two leptons of different flavors. 
The $\PSgn_\PGt$ is produced from the annihilation of two down quarks via the $\lambda'_{311}$ coupling, and then 
decays via the $\lambda$ couplings. Middle: Resonant production of a \PZpr boson with subsequent
decay into two leptons of different flavors. Right: Production of quantum black holes in a model with extra dimensions that involves subsequent
decay into two leptons of different flavors.
} 
\label{Fig:feyn}
\end{figure}

Theories that invoke extra spatial dimensions can lower the fundamental Planck scale to the TeV region.
Such theories also provide the
possibility of producing microscopic QBHs~\cite{Calmet:2008dg,Meade:2007sz} at the LHC. In contrast to semiclassical thermal
black holes that can decay to high-multiplicity final states, QBHs are nonthermal objects, expected to decay predominantly to
pairs of particles. We consider the production of spin-0, colorless, neutral QBHs in a model with LFV~\cite{Gingrich},
in which the cross section for QBH production depends on the threshold mass $m_{\text{th}}$ in $n$ additional
spatial dimensions. The $n = 1$ possibility corresponds to the Randall--Sundrum (RS) brane-world model~\cite{Randall:1999ee}, and $n>1$
corresponds to the Arkani-Hamed--Dimopoulos--Dvali (ADD) model~\cite{ArkaniHamed:1998rs}. While the resonant $\PSgn_\PGt$ and
\PZpr signals are expected to generate narrow peaks in the invariant mass spectrum of the lepton pair, the distribution of the QBH signal is
characterized by a sharp edge at the threshold of QBH production, followed by a monotonic decrease at larger masses. Feynman diagrams for all these three models are shown 
in Fig.~\ref{Fig:feyn}.

Similar searches in LFV dilepton mass spectra have been carried out by the CDF~\cite{Aaltonen:2010fv} and D0~\cite{Abazov:2010km}
experiments at the Fermilab Tevatron in proton-antiproton collisions at $\sqrt{s}=1.96\TeV$ and by the ATLAS
and CMS experiments at the LHC in $\Pp\Pp$ collisions at $\sqrt{s}=8\TeV$~\cite{Aad:2015pfa,Khachatryan:2016ovq} and
13\TeV~\cite{TheATLAScollaboration:2015ucd,Sirunyan:2018zhy,ATLAS:2018mrn}.

This paper is structured as follows. The CMS detector is briefly described in Section~\ref{sec_det}. 
A description of the collision data and the simulated event samples used in the analysis is given in Section~\ref{sec_datamc}. 
The event reconstruction and selection are described in Section~\ref{sec_evtsel} and the estimation of SM backgrounds is discussed in Section~\ref{sec_bkg}. 
Systematic uncertainties are described in Section~\ref{sec_syst}, followed by the results and their statistical interpretation
in Section~\ref{sec_result}. The paper is summarized in Section~\ref{sec_summary}. 

Tabulated results are provided in the HEPData record for this analysis~\cite{hepdata}.

\section{The CMS detector}
\label{sec_det}
The central feature of the CMS apparatus is a superconducting solenoid of 6\unit{m} internal diameter, 
providing a magnetic field of 3.8\unit{T}. Within the solenoid volume are a silicon pixel and strip tracker, 
a lead tungstate crystal electromagnetic calorimeter (ECAL), and a brass and scintillator 
hadron calorimeter (HCAL), each composed of a barrel and two endcap sections. 
Forward calorimeters extend the pseudorapidity ($\eta$) coverage 
provided by the barrel and endcap detectors. Muons are detected in gas-ionization chambers embedded in 
the steel flux-return yoke outside the solenoid. Muons are measured in the pseudorapidity range $\abs{\eta} < 2.4$, with detection planes made using three technologies: drift tubes, cathode strip chambers, and resistive-plate chambers.
Isolated particles of transverse momentum (\pt) of 100\GeV emitted within the pseudorapidity range $\abs{\eta} < 1.4$ have track resolutions of 2.8\% in \pt and 10 (30)\mum in 
the transverse (longitudinal) impact 
parameter \cite{TRK-11-001}. 

Events of interest are selected using a two-tiered trigger system. The first level (L1), composed of custom hardware processors, uses information from the calorimeters and muon detectors to select events at a rate of around 100\unit{kHz} within a fixed latency of about 4\mus~\cite{CMS:2020cmk}. The second level, known as the high-level trigger (HLT), consists of a farm of processors running a version of the full event reconstruction software optimized for fast processing, and reduces the event rate to around 1\unit{kHz} before data storage~\cite{CMS:2016ngn}.

The single-muon trigger efficiency exceeds 90\% over the full $\eta$ acceptance, and the efficiency to reconstruct and identify muons is greater 
than 96\%. Matching muons to tracks measured in the silicon tracker results in a relative \pt resolution, for muons with \pt up to 100\GeV, of 1\% in 
the barrel and 3\% in the endcaps. The \pt resolution in the barrel is better than 7\% for muons with \pt up to 1\TeV~\cite{muon_paper}.

The single isolated electron trigger efficiency exceeds 80\% over the full $\eta$ acceptance of $\abs{\eta} < 2.5$, and the efficiency to reconstruct electrons is greater than 95\% for electrons with transverse momentum larger than 20\GeV~\cite{CMS:2020uim}. 
The electron momentum is estimated by combining the energy measurement in the ECAL with the momentum measurement in the tracker. 
The momentum resolution for electrons with $\pt \approx 45\GeV$ from $\PZ \to \Pe \Pe$ decays ranges from 1.7 to 4.5\%. It is generally better in the 
barrel region than in the endcaps, and also depends on the bremsstrahlung energy emitted by the electron as it traverses the material in front 
of the ECAL~\cite{Khachatryan:2015hwa}.

A more detailed description of the CMS detector, together with a definition of the coordinate system used and the relevant kinematic variables, can be found 
in Ref.~\cite{Chatrchyan:2008zzk}.

\section{Collision data and simulated events}
\label{sec_datamc}
The data sample used for this analysis was collected during 2016--2018 $\Pp\Pp$ operation at a center-of-mass energy of 13\TeV.
After applying data quality
requirements, such as requiring minimum fractions of active detector channels, the total integrated luminosity is 138\fbinv.

Simulated samples of signal and background events are produced with several event generators. 
The RPV SUSY $\PSgn_\PGt$, \PZpr, and QBH signal events are generated at leading order (LO) precision, using
the \CALCHEP~3.6~\cite{Belyaev:2012qa}, \PYTHIA~8.203~\cite{Sjostrand:2014zea}, and \textsc{qbh}~3.0~\cite{Gingrich} Monte Carlo (MC) generators,
respectively. The width of the \PZpr signal is taken as
3\% of its mass, similar to that of the \PZ boson.
The RPV $\PSgn_\PGt$ and QBH signals
are generated using the CTEQ6L~\cite{Pumplin:2002vw} parton distribution functions (PDF), while the \PZpr boson signals are simulated using the
NNPDF 3.0 PDF set in 2016 and the 3.1 PDF set~\cite{Ball:2014uwa} in 2017--2018, respectively. 
The LO RPV SUSY $\PSgn_\PGt$ signal event yield is normalized to a next-to-leading order (NLO)
calculation of the production cross section~\cite{Dreiner:2006sv}; in this calculation the factorization and renormalization scales are set to the mass of
the $\PSgn_\PGt$. 
The \POWHEG v2.0 event generator~\cite{Nason:2004rx,Frixione:2007vw,Alioli:2010xd,Frixione:2007nw} is used to simulate both top quark pair 
($\PQt\PAQt$) production, the dominant background over most of the dilepton mass region, and single top quark production. Diboson production, a 
significant background in the high mass region, is simulated at LO using the \PYTHIA generator.

The \MGvATNLO v2.2.2 generator~\cite{Alwall:2014hca} is used to simulate Drell--Yan+jets production simulated at NLO
with the FxFx jet matching and merging~\cite{Frederix:2012ps}.
The cross sections used to normalize the contribution of these backgrounds are calculated
at next-to-next-to-leading order for $\PW\PW$, $\PZ\PZ$, single top quark, and $\PQt\PAQt$ processes, and at NLO precision for $\PW\PZ$ and Drell--Yan events.
The \POWHEG and \MGvATNLO generators are interfaced with \PYTHIA for parton showering, fragmentation, and decays. 
The \PYTHIA parameters for the underlying event description are set
to the CUETP8M1 (CP5) tune~\cite{Khachatryan:2015pea} in 2016 (2017--2018) simulated samples. 

The generated events are processed through a full simulation of
the CMS detector, based on \GEANTfour~\cite{Agostinelli:2002hh,Allison:2006ve,ALLISON2016186}. The simulated events
incorporate additional $\Pp\Pp$ interactions within the same or nearby bunch crossings, termed pileup, that are weighted
to match the measured distribution of the number of interactions per bunch crossing in data. The simulated event samples
are normalized to the integrated luminosity of the data. The products of the total acceptance and
efficiency for the three signal models in this analysis are determined through MC simulation. The trigger and
object reconstruction efficiencies are corrected to the values measured in data.

\section{Event reconstruction and selection}
\label{sec_evtsel}

The global event reconstruction is performed using a particle-flow algorithm~\cite{CMS-PRF-14-001}, 
which aims to reconstruct and identify each individual particle with an optimized combination of all subdetector
information. In this process, the identification of the particle type (photon, electron, muon, and charged or neutral hadron) plays an 
important role in the determination of the particle's direction and energy. 

The primary vertex (PV) is taken to be the vertex corresponding to the hardest scattering in the event, evaluated using tracking information alone, 
as described in Section 9.4.1 of Ref.~\cite{CMS-TDR-15-02}.

To reconstruct an electron candidate, energy deposits in the ECAL are first combined into
clusters, assuming that each cluster represents a single particle. The clusters are then combined
in a way consistent with bremsstrahlung emission, to produce a single ``supercluster'', which
represents an electron or photon. These superclusters are used to seed tracking algorithms,
and if a resulting track is found, it is assigned to the supercluster to form an electron candidate. 

To reconstruct a muon candidate, hits are first fitted separately to 
trajectories in the inner tracker detector and in the outer-muon system. The two trajectories are then combined in a
global-muon track hypothesis. 

Hadronic $\PGt$ lepton decays ($\tauh$) are reconstructed from jets, using the hadrons-plus-strips algorithm~\cite{CMS:2018jrd}, which combines 1 or 3 tracks with energy deposits in the calorimeters, to identify the $\PGt$ lepton decay modes. 
Neutral pions are reconstructed as strips with dynamic size in $\eta$-$\phi$, where $\phi$ is the azimuthal angle in radians,  
primarily from reconstructed photons, but also from reconstructed electrons which originate due to conversion of photons. 
The strip size varies as a function of the $\pt$ of the electron or photon candidate.

To distinguish genuine $\tauh$ decays from jets originating from the hadronization of quarks or gluons, and from electrons or muons, 
the \textsc{DeepTau} algorithm is used~\cite{TAU-20-001}. 
Information from all individual reconstructed particles near the $\tauh$ axis is combined with properties of the $\tauh$ candidate in the event 
to yield separate discriminants for jet, electron, and muon backgrounds.
The probability that a jet is misidentified as $\tauh$ by the \textsc{DeepTau} algorithm depends on the $\pt$ and quark flavor of the jet. In simulated events of $\PW$ boson production in association with jets, this probability has been estimated to be $0.43\%$ for a $\tauh$ identification efficiency of $70\%$. 
The probability that an electron (muon) is misidentified as a $\PGt$ is $2.60\,(0.03)\%$ for a $\tauh$ identification efficiency of $80\,({>}99)\%$. 

\subsection{Selection of events for the \texorpdfstring{$\Pe\PGm$}{emu} final state}

For the $\Pe\PGm$ selection, at least one prompt,
isolated electron and at least one prompt, isolated muon are required in the event. This minimal selection
facilitates a reinterpretation of the results in terms of models with more complex signal
topologies than a single $\Pe\PGm$ pair.

Events that satisfy single electromagnetic-cluster or single-muon triggers, with respective \pt thresholds of 175 and 50\GeV for photons and muons, 
are selected. In 2018, the \pt threshold of the photon trigger was raised to 200\GeV. 
Electromagnetic energy deposited by an electron in the
calorimeter activates the photon trigger, and the photon trigger is 
therefore as efficient as the corresponding electron trigger, while its weaker isolation requirements
yield an event sample that can also be used in sideband analyses to estimate the background
to the signal. Moreover, using only the muon trigger yields about 90\% signal efficiency in masses above 1\TeV. Using the photon trigger together 
with the muon trigger ensures that the signal efficiency is very close to 100\% in the high mass region.

The electron candidate must pass the dedicated selection criteria developed for high-energy electrons~\cite{CMS:2020uim}, 
which require the energy deposition in the ECAL to be consistent with that of an electron.
The electron candidates are required to have $\pt > 35\GeV$ and $\abs{\eta} < 2.5$.
The energy sum in the HCAL within a cone defined by $\DR = \sqrt{\smash[b]{(\Delta\eta)^2+(\Delta\phi)^2}} = 0.15$ centered around the electron
candidate must be ${<}5\%$ of its energy after it is corrected for jet activity unrelated to the primary interaction. 
The electron candidate must have a prompt track in the $\eta$-$\phi$ plane that
has at most one missing hit in the pixel tracker and that is well-matched between the tracker and ECAL. The high-energy electron selection also
requires electrons to be isolated. 
The scalar-\pt sum of tracks
within a cone of radius $\DR = 0.3$ around the candidate's direction, excluding the candidate
track, is required to be ${<}5\GeV$. Moreover, the \pt sum of energy depositions in the calorimeters within the same 
cone, excluding the ECAL supercluster, is required to be ${<}3\%$ of the \pt of the candidate~\cite{CMS:2020uim}.
The efficiency of the electron identification requirements, measured in data, is approximately $90\%$ for the highly energetic electrons that are relevant for this search. 

Muon candidates are required to have $\pt > 53 \GeV$ and $\abs{\eta} < 2.4$.
The muon candidate must pass the high-\pt muon identification criteria, which require that 
the transverse and longitudinal impact parameters of muon candidates relative to the primary 
vertex must be ${<}0.2$ and ${<}0.5\unit{cm}$, respectively. 
The track of the muon
candidate must have at least one hit in the pixel detector and hits in at least six silicon-strip
layers, and must contain matched segments in at least two muon detector planes. 
To suppress
backgrounds arising from muons within jets, the scalar-\pt sum of all other tracks in the tracker
within a cone of $\DR = 0.3$ around the muon candidate track is required to be $<$10\% of the
 \pt of the muon candidate. The relative uncertainty in the \pt of the muon track is required to be ${<}30\%$.
The efficiency of the muon identification and isolation requirements, measured in data, is about $90\%$ for muons with $\pt > 50 \GeV$.

To prevent a loss in signal efficiency resulting from the misidentification of the sign of the electron or muon
charge at large \pt, the electron and muon are not required to have opposite charges. Since
highly energetic muons can produce bremsstrahlung in the ECAL along the direction of the
inner-muon trajectory, such muons can be misidentified as electrons. An electron candidate
is therefore rejected if there is a muon candidate track with $\pt > 5 \GeV$ within 
$\DR < 0.1$ of the electron candidate track. Only one $\Pe\PGm$ pair is considered per event.
When there is more than one $\Pe\PGm$ candidate, the pair with the highest invariant mass is selected
for the analysis. 
The statistical interpretation is conducted by comparing the shape of the observed invariant $\Pe\PGm$ mass distribution with those expected
for the signal and background.

\subsection{Selection of events for the \texorpdfstring{$\Pe\PGt$}{etau} and \texorpdfstring{$\PGm\PGt$}{mutau} final states}

Events are required to have at least one prompt, isolated light lepton (electron or muon) and at least one prompt, isolated $\tauh$.

Events that satisfy single-electron triggers with thresholds of 27, 35, and 32\GeV in 2016, 2017, and 2018, respectively, are selected for the $\Pe\PGt$ channel.
To recover efficiency in the high mass region, events that satisfy single electromagnetic cluster triggers are also selected for this channel.
The electron candidate must pass the same high-energy electron selection used for the $\Pe\PGm$ channel, but with an increased \pt threshold of 50\GeV.

The single-muon triggers used in the $\Pe\PGm$ channel are also used to collect the data samples in the $\PGm\PGt$ channel.
The muon candidate must pass the same high-\pt muon identification criteria used for the $\Pe\PGm$ channel.

The $\tauh$ candidate having the largest transverse momentum must have $\pt>50\GeV$ and $\abs{\eta} < 2.3$.
To reduce the rate of jets, electrons, and muons misidentified as $\tauh$, the $\tauh$ candidate is required to satisfy the tight, 
loose, and tight working points of the respective \textsc{DeepTau} discriminator~\cite{TAU-20-001}.

The low transverse mass ($\mT$) region is dominated by misidentified $\tauh$ events, which have 
no genuine neutrinos, so the missing transverse momentum is small.
Thus, a
requirement of $\mT > 120\GeV$ is applied which helps to reject misidentified $\tauh$ background, where \mT is defined as:
\begin{linenomath}
\begin{equation}
\mT = \sqrt{2 \pt^{\ell} \, \ptmiss [1-\cos\Delta\phi(\ptvec^{\ell},\ptvecmiss)]},
\end{equation}
\end{linenomath}
where $\ell$ denotes the light lepton, \ie, the electron in the $\Pe\PGt$ final state or the muon in the $\PGm\PGt$ final state, \ptvecmiss is the 
missing transverse momentum vector in the event, and $\Delta\phi$ is the difference in the azimuthal angle between $\ptvec^{\ell}$ and $\ptvecmiss$.

A muon veto is applied in the $\Pe\PGt$ final state to remove overlap with the $\PGm\PGt$ and $\Pe\PGm$ final states. 
This veto rejects events if they contain any isolated muon with $\pt > 35\GeV$, $\abs{\eta} < 2.4$, passing the high-\pt muon identification criteria and having
tracker-based isolation $<$0.15. Events with a well-separated electron pair are also rejected. 
A well-separated electron pair is defined as two electrons with $\DR(\Pe,\Pe)>0.5$, each of which has $\pt > 10\GeV$, $\abs{\eta} < 2.5$, and passes a very 
loose working point (${>}95\%$ efficiency) of the electron identification criteria.

In the $\PGm\PGt$ channel, in order to avoid overlap with the $\Pe\PGt$ and $\Pe\PGm$ channels, events are vetoed if they contain an electron with $\pt > 35\GeV$ that 
passes the high-energy electron selection criteria.
Events with a well-separated muon pair are also rejected. A well-separated muon pair is defined as two muons passing the high-\pt identification criteria, with $\pt > 10\GeV$,
$\abs{\eta} < 2.4$, tracker-based isolation $< 0.15$, and with $\DR (\PGm, \PGm) > 0.2$.

If there is more than one $\Pe\PGt$ or $\PGm\PGt$ pair in an event, then the pair with highest invariant mass is chosen. 
The leptons forming the pair are not required to have opposite electric charges.
The statistical interpretations are conducted by comparing the shapes of the observed collinear mass distributions with those expected for the signals 
and backgrounds. 
For an $\Pe\PGt$ pair or a $\PGm\PGt$ pair, we use the collinear mass, which provides an estimate of the mass 
of the new resonance or QBH based on their observed decay products. 
It is justified by the observation that, 
since the mass scale of the signal is orders of magnitude higher than that of the $\PGt$ lepton, 
the $\PGt$ lepton decay products are 
highly Lorentz boosted in the direction of the $\PGt$ candidate. The neutrino momentum can be approximated to have the
same direction as the other, visible decay products of the $\PGt$ lepton, so the component of \ptvecmiss in the direction of the visible 
$\PGt$ lepton decay products is used to estimate the neutrino \pt.  
The variable $x_{\PGt}^{\text{vis}}$, which is the fraction of energy carried by the visible decay products of the $\PGt$, is defined as $x_{\PGt}^{\text{vis}}  = \pt^{\text{vis}} / ( \pt^{\text{vis}} + p_{\text{T, coll}}^{\text{miss}} )$, where $p_{\text{T, coll}}^{\text{miss}}$ is the part of the \ptmiss that is collinear with the $\tauh$ \pt.
The collinear mass $m_{\text{coll}}$ can then be derived from
the visible mass $m_{\text{vis}}$ of the $\Pe\PGt$ or $\PGm\PGt$ system to be
$m_{\text{coll}} = m_{\text{vis}} / \sqrt{\smash[b]{ x_{\PGt}^{\text{vis}}}  } $,
where $m_{\text{vis}}$ is the invariant mass of the visible $\PGt$ decay products.

\section{Background estimation}
\label{sec_bkg}
The SM background in the LFV dilepton search includes several processes that produce a final state with two different-flavor charged leptons. For all channels, the dominant background contributions originate from $\PQt\PAQt$ production.   
Other less significant backgrounds originate from diboson ($\PW\PW$, $\PW\PZ$, and $\PZ\PZ$), $\PZ \to \ell^+ \ell^-$ ($\ell = \Pe$, $\PGm$, $\PGt$), $\PW\PGg$, and single top quark production processes. 
All these backgrounds are estimated from MC simulation. 
For the $\PZ \to \PGt \PGt$ process, all decay modes of $\PGt$ lepton have been considered.                            
Multijet and $\PW$+jets processes also contribute due to the misidentification
of jets as leptons. These backgrounds are estimated from data.  
         
For the $\Pe\PGm$ channel, to determine the contributions from $\PW$+jets and multijet processes to the $m_{\Pe\PGm}$ distribution, a control sample in data is defined using jet-to-electron misidentification rate ($F_\Pe$).
This rate is obtained from data, by selecting electron candidates with relaxed selection requirements, using events collected with a single electromagnetic-cluster trigger. Data sidebands obtained by inverting the selection requirements on either the electron isolation or shower shape variables are used to evaluate the contribution of jets passing the full electron selection in the control sample~\cite{CMS:2014lcz}. The jet-to-electron misidentification rate is then defined as the number of jets passing the full electron selection divided by the number of jet candidates in the sample. 
The rate is quantified in bins of misidentified electron \pt and pseudorapidity. The measured jet-to-electron rate is then 
used to estimate the $\PW$+jets and multijet contributions to $\Pe\PGm$ using data containing muons that pass
the single-muon trigger and the full muon selection, and electron candidates
satisfying relaxed selection requirements, but failing the full electron selection. Each event
is weighted by the factor $F_\Pe/(1-F_\Pe)$ to determine the overall contribution from the jet backgrounds. 
A correction is applied to account for the fraction of events in the control sample that have genuine leptons, estimated from simulation.
The background originating from jets misidentified as electrons is about 10\% of the total background.

Background from jets mimicking muons is estimated in a similar way, 
where identification criteria are loosened for muons in order to create a control sample enriched with jet candidates that are misidentified as muons. Then the jet-to-muon misidentification rate $F_{\PGm}$ is defined as the number of jets passing the full muon selection divided by the number
of jet candidates in the sample.
The rate is quantified in bins of misidentified muon $\pt$ and $\eta$ as well. Events in data that have one well-identified electron and 
one candidate satisfying loose requirements of muon identification 
are selected, with the weight factor $F_{\PGm}/(1-F_{\PGm})$ applied to estimate the jet-to-muon misidentification contribution in the signal region.
The background originating from jets misidentified as muons is about 2\% of the total background.
Events with both electron and muon misidentified from jets are not considered since their contribution is expected to be small.

In the $\PGm\PGt$ and $\Pe\PGt$ channels, the most significant background after the $\PQt\PAQt$ and $\PW\PW$ processes comes
from $\PW$+jets and multijet processes, where jets are misidentified as $\tauh$ candidates.                       
This background is estimated using collision data, in a control sample with the same selection as the signal region, except that the \mT
requirement is inverted to $\mT < 120\GeV$.
From this low-\mT control sample, two subsamples are constructed to compute the probability for an accompanying jet to be misidentified as 
a $\tauh$:
\begin{itemize}
\item Subsample A requires $\tauh$ candidates to fail the tight antijet discriminator working point but pass the loose working point.
\item Subsample B requires $\tauh$ to pass the tight working point.
\end{itemize} 

A factor $F_{\PGt}$ is calculated from these samples, defined as the ratio of the number of events in subsamples B and A. This factor is calculated as a function of the 
$\pt$ of the $\tauh$ candidate, the ratio of the $\pt$ of the $\tauh$ candidate to the $\pt$ of its parent jet, and the pseudorapidity of the $\tauh$. The number of events expected from misidentified $\tauh$ in the signal region is estimated from a control sample fulfilling the same criteria as events in the signal region, except that the $\tauh$ candidates pass the looser working point but fail the tight working point of the antijet discriminator. Each event is weighted by the factor $F_{\PGt}$ to determine its effective contribution. A correction is applied to account for the fraction of events in the control sample that have genuine $\tauh$ candidates, estimated from simulation.

\section{Systematic uncertainties}
\label{sec_syst}
Various effects impact the shape and the normalization of the invariant or collinear mass distributions and lead to 
systematic uncertainties in the signal and background estimates.
The uncertainty in the modeling of the invariant or collinear mass distributions reflects three
types of systematic effects. 

The first type includes those that affect both the shape and normalization of the mass distributions. For all channels, the
dominant uncertainties arise from the leading $\PQt\PAQt$ and subleading $\PW\PW$ backgrounds. 
They result in a 30--50\% uncertainty in the number of $\PQt\PAQt$ and $\PW\PW$ events at a dilepton mass scale of 2\TeV.
The uncertainty in the $\PW\PW$ background is estimated from the envelope of the resummed next-to-next-to-leading logarithmic calculation of the soft-gluon contributions to the cross section at NLO, as presented in Ref.~\cite{Pecjak:2016nee}, 
using changes in the renormalization and factorization scales by factors of 2 and 0.5.
Similarly, the uncertainty in the $\PQt\PAQt$ background is estimated by considering the variations in PDF and factorization 
scales, as discussed in Ref.~\cite{Czakon:2017wor}. Other contributions include the uncertainty in the muon momentum scale, which is approximately 1--2\% for 1\TeV muons and depends on their $\eta$ and $\phi$~\cite{CMS:2021ctt,CMS:2019ied}. 
The uncertainty in the muon efficiency ($0.5$\% for 1\TeV muons) is considered in the $\Pe\PGm$ and $\PGm\PGt$ channels~\cite{CMS:2019ied},
and the uncertainty in the electron efficiency (varies between 1--5\%, depending on \pt and $\eta$ of the electrons) is considered in 
the $\Pe\PGm$ and $\Pe\PGt$ channels~\cite{CMS:2020uim}. In the $\PGt$ channels, the uncertainties in the $\tauh$ identification (5\% for 1\TeV $\PGt$ leptons) and $\tauh$ energy scale (1.5--4.0\% for $\pt (\tauh) > 100\GeV$, depending on the decay mode) are considered~\cite{CMS:2018jrd}. Uncertainties in the electron \pt scale and resolution, the muon \pt scale and resolution, and the pileup rate are also taken into account, but they have 
negligible impact on the total background. The uncertainties in the determination of the trigger efficiencies have
a very small impact on the expected event yields. The uncertainty associated with the choice of the PDF in the simulation is evaluated according to the PDF4LHC prescription~\cite{Butterworth:2015oua}.

The jet energy scale is determined with an uncertainty amounting to a few percent, depending on the jet \pt and $\eta$,
using the \pt-balance method. This correction is applied to $\PZggx \to \Pe\Pe$, $\PZggx \to \PGm\PGm$, $\PGg$+jets, dijet, and multijet events~\cite{Khachatryan:2016kdb}.
The resulting effect on signal and background expectations is evaluated by varying the energies of jets in simulated events within their uncertainties,
recalculating all kinematic observables, such as \ptvecmiss, and reapplying the event selection criteria. The effects of uncertainties in the energy scale of the unclustered particles and jet energy resolution are evaluated in a similar way. These systematic uncertainties 
affect the shapes as well as the normalizations of the collinear mass distributions.

Uncertainties of the second type directly influence only the normalization of the 
mass distribution. A systematic uncertainty of 1.6\% in the integrated luminosity~\cite{CMS:2021xjt,CMS:2018elu,CMS:2019jhq} is taken for the
backgrounds and signals. Among the uncertainties in the cross sections used for the normalization of various simulated 
backgrounds, the 5\% uncertainty in the $\PQt\PAQt$ background is the 
most important. A systematic uncertainty of 50\% is applied to the estimate of the misidentified jet background
derived from data in all three channels. For $\tauh$ final states, this uncertainty is obtained by using the misidentification probability $F_{\PGt}$ derived from an independent control sample of \PZ($\to\PGm\PGm$)+jets events for the estimation of misidentified jet background. 
It is found, especially at high masses (${\ge}1.5\TeV$), that the collinear mass distributions obtained in the signal region using the $F_{\PGt}$ calculated 
from the two independent 
control samples agree within 50\% and are consistent within the statistical uncertainties.
Therefore, an overall 50\% systematic uncertainty is assigned to the estimation of this background.

Uncertainties of the third type are associated with 
limited sizes of event samples in the MC simulation of background processes~\cite{Barlow:1993dm}. In contrast to other uncertainties, they are not correlated
between bins of the invariant mass distribution.

Taking all systematic uncertainties into account, the resulting relative uncertainty 
in the background increases with mass. This relative increase does not significantly affect the sensitivity at large mass values, where the
expected number of events from SM processes becomes negligible. All the relevant uncertainties are
also taken into account in the extraction of various signals.

All uncertainties are considered to be correlated across the different data-taking years, with the exception of the $\tauh$ object-related uncertainties and the uncertainty in the unclustered energy, which are derived from statistically independent sources. No correlation between the different final states is considered, and the analysis results are presented independently for each of the final states.
 
\section{Results and their interpretations}
\label{sec_result}

\begin{figure}[p]
\centering
\includegraphics[width=.49\textwidth]{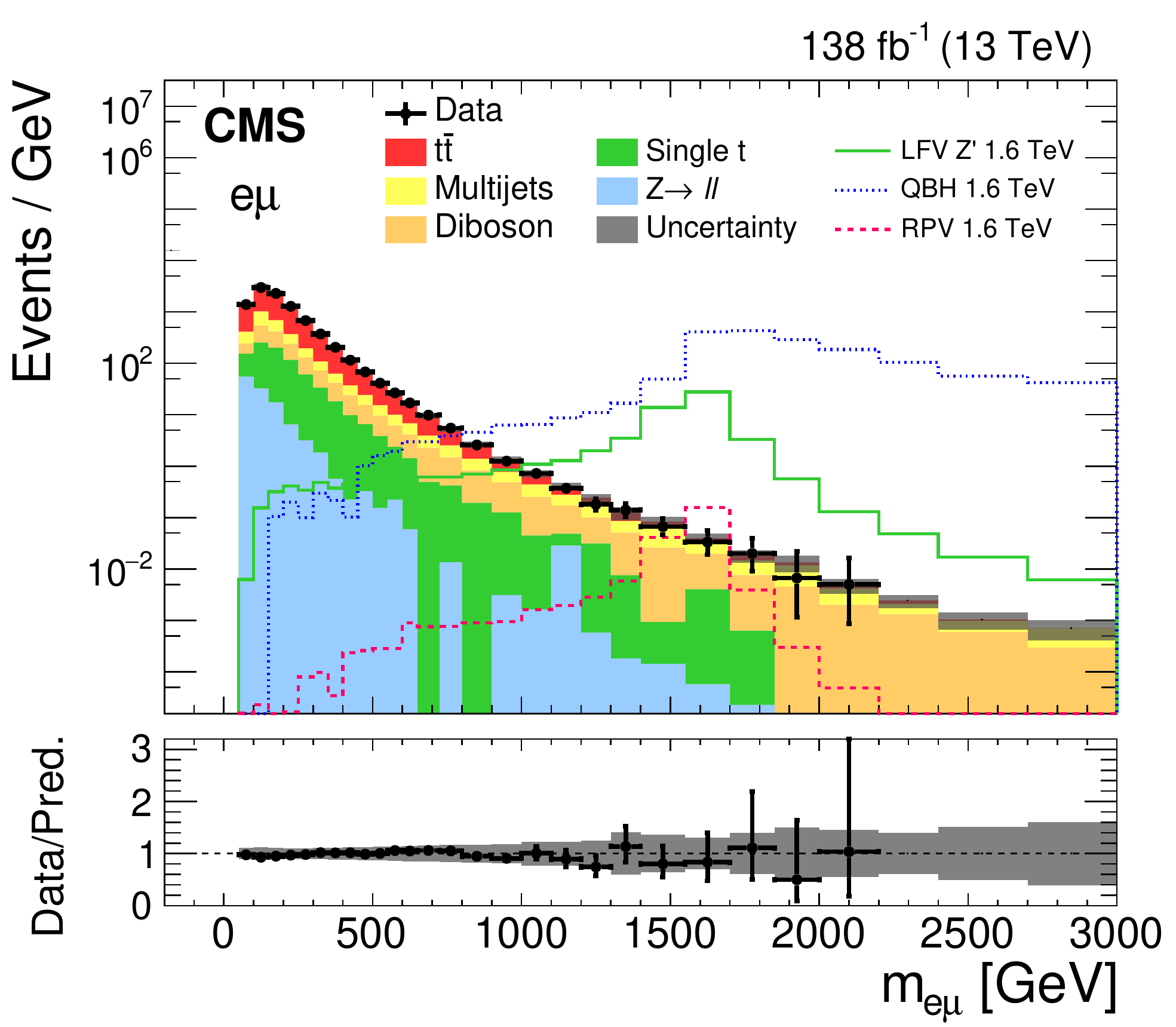}\\
\includegraphics[width=.49\textwidth]{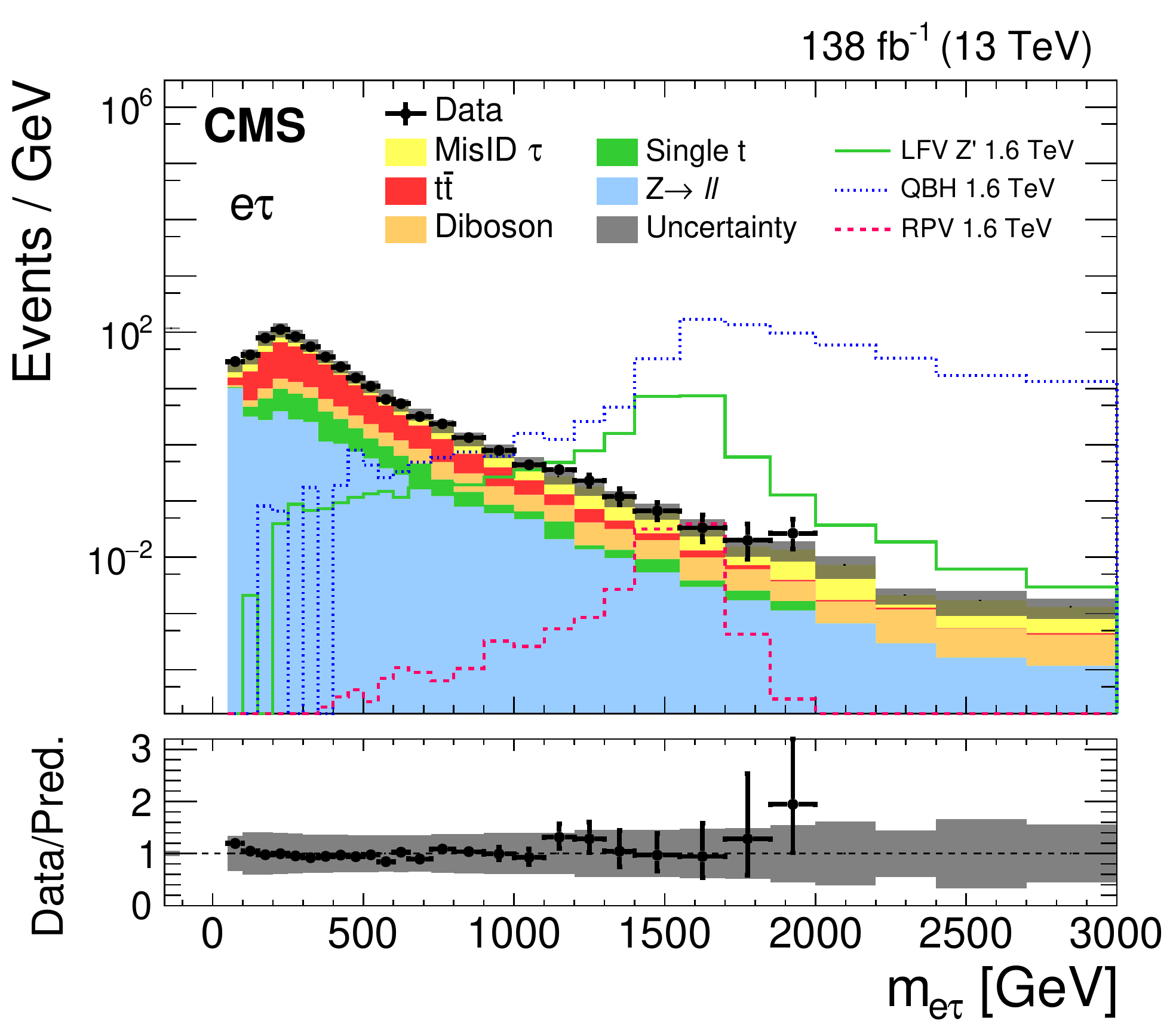}
\includegraphics[width=.49\textwidth]{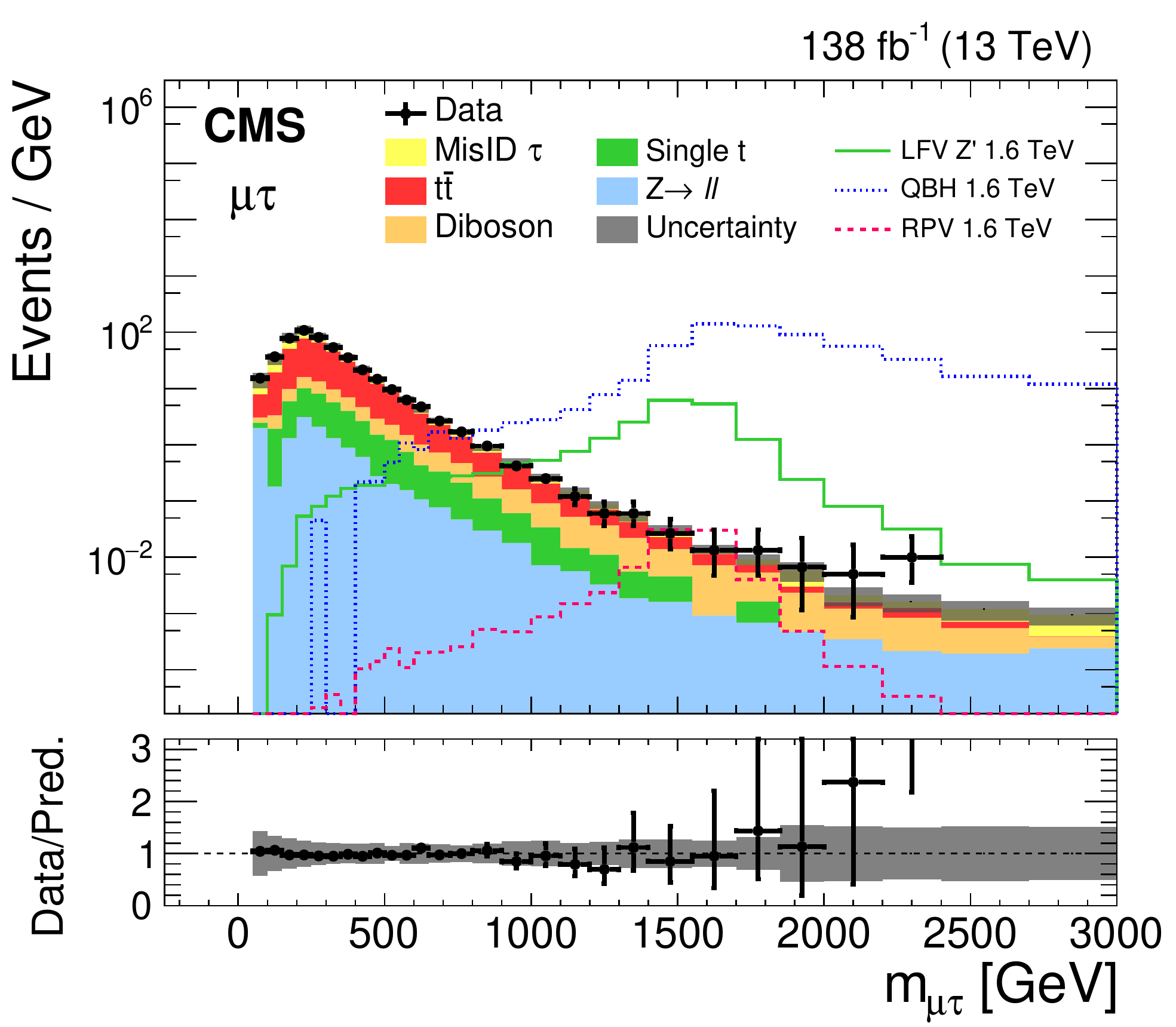}

\caption{Invariant mass distributions for the $\Pe\PGm$ channel (upper), and collinear mass distributions for the $\Pe \PGt$ (lower left) and $\PGm \PGt$ (lower right) channels. In addition to the observed data (black points) and the SM prediction  
(filled histograms), the expected signal distributions for three models are shown: the RPV SUSY model with $\lambda = \lambda' = 0.01$ and $\PGt$ sneutrino mass of 1.6\TeV, 
LFV \PZpr ($\mathcal{B} = 0.1$) boson with a mass of 1.6\TeV, and the QBH signal expectation for $n=4$ and a threshold mass of 1.6\TeV. The bottom panel of each plot shows the ratio of data and SM prediction. The bin width gradually 
increases with mass.} 
\label{Fig:m_tot}
\end{figure}

The mass distributions in the $\Pe\PGm$, $\Pe\PGt$, and $\PGm\PGt$ channels, shown in Fig.~\ref{Fig:m_tot}, do not exhibit 
significant deviations from the expected SM background. 
The last bins with data events show minor statistical fluctuations in the $\Pe \PGt$ and $\PGm \PGt$ channels, which are consistent 
with the SM expectations within 2 standard deviations, as shown later in the limit plots.
The expected mass distributions of signal and backgrounds are taken from
simulation, with the exception of backgrounds with jets misidentified
as leptons, which are estimated using collision data, as discussed in previous sections.
Upper limits on the products of the production cross section $\sigma$ and branching fraction $\mathcal{B}$ are determined 
using a Bayesian binned-likelihood method~\cite{CowanPDGStat,ATLAS:2011tau}
 with a uniform positive prior probability density for the signal cross section. The nuisance parameters associated with the 
systematic uncertainties are modeled via log-normal distributions for uncertainties in the normalization. Uncertainties in the shape of the distributions are 
modeled via ``template morphing'' techniques~\cite{Baak:2014fta}. A Markov Chain MC method~\cite{Allanach:2007qj} is used for the integration. All limits presented here are at 95\% confidence level (\CL). 

\begin{figure}[p]
\centering
\includegraphics[width=.49\textwidth]{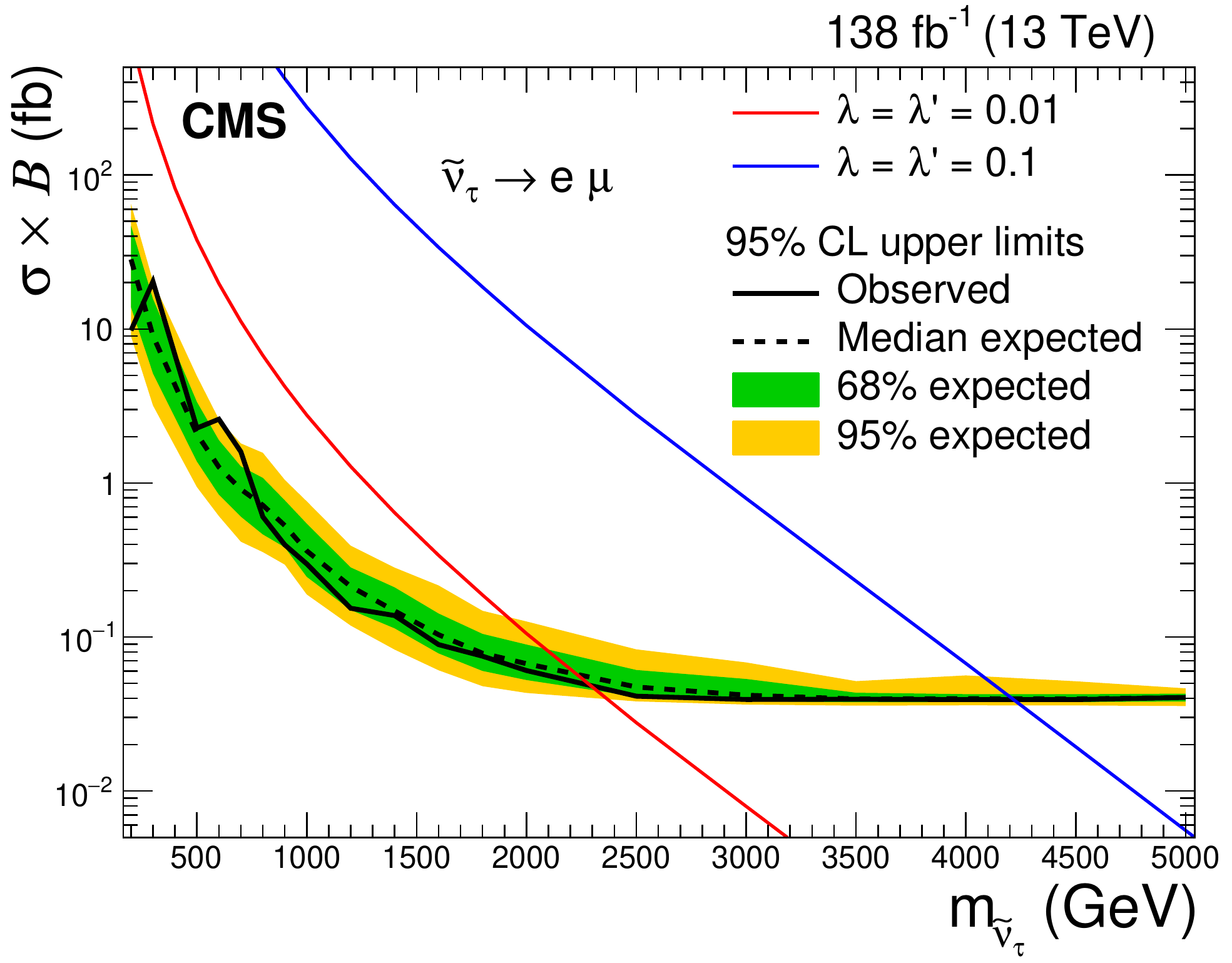}\\
\includegraphics[width=.49\textwidth]{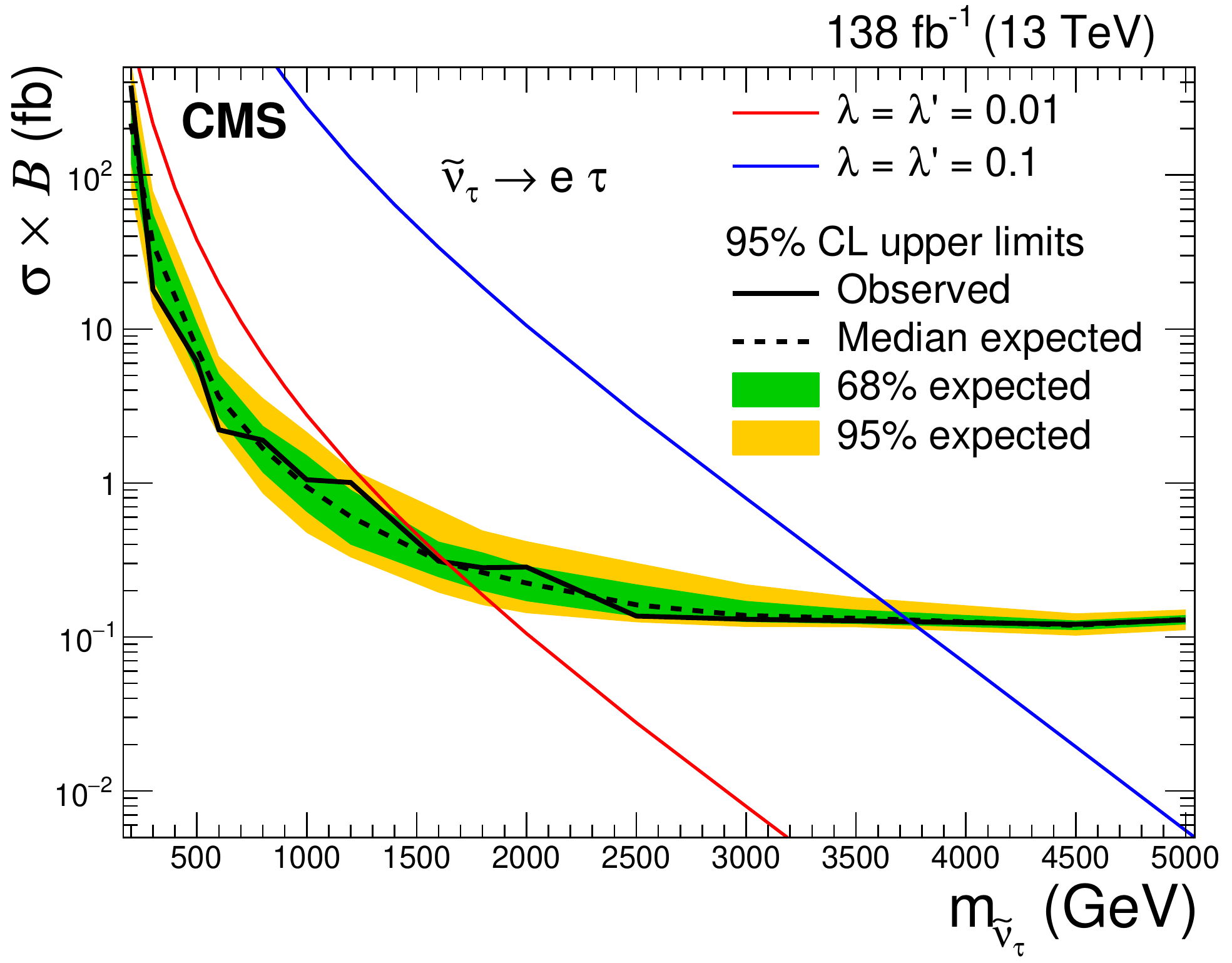}
\includegraphics[width=.49\textwidth]{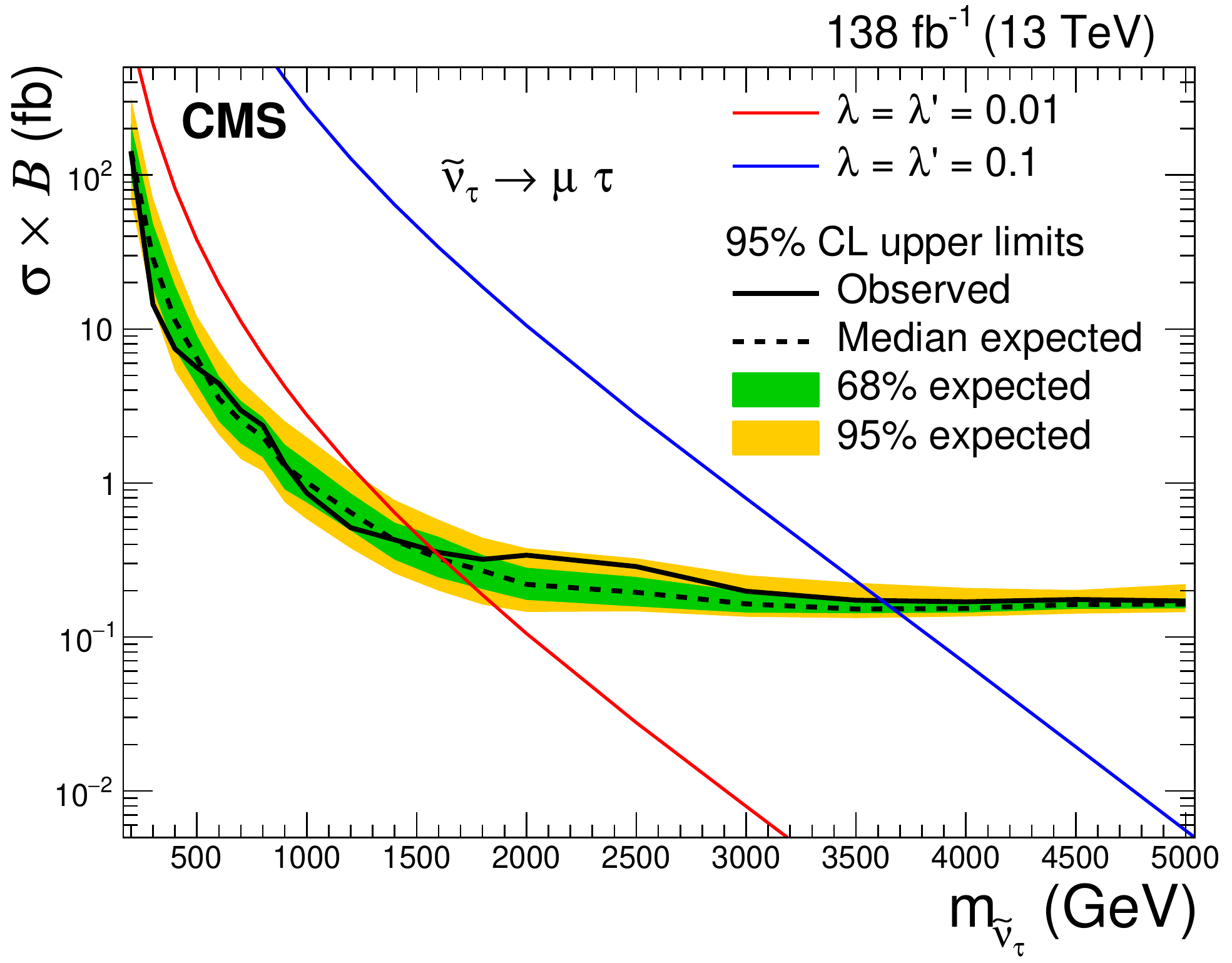}
\caption{
Expected (black dashed line) 
and observed (black solid line) 
95\% \CL  upper limits on the product of the cross section and the branching fraction as a function of the $\PGt$ sneutrino mass in an RPV SUSY model for the $\Pe\PGm$ (upper), $\Pe\PGt$ (lower left), and $\PGm\PGt$ (lower right) channels.
The shaded bands represent 
68\% and 95\% uncertainties in the expected limits. The red and blue solid lines show the predicted product of the cross section and the branching fraction as a function of the tau sneutrino mass for two different values of the couplings.
}
\label{Fig:lim_rpv}
\end{figure}

\begin{figure}[p]
\centering
\includegraphics[width=.49\textwidth]{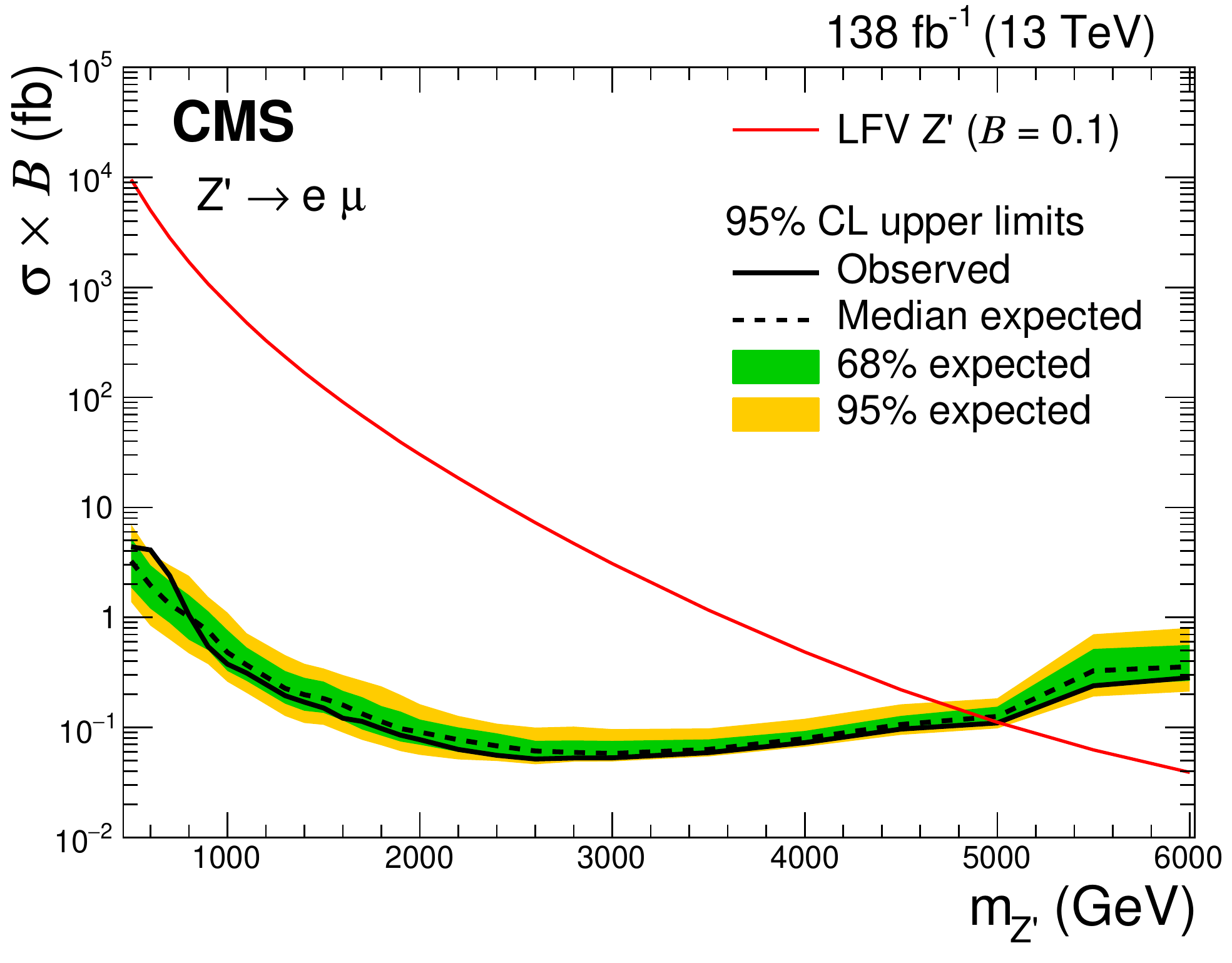} \\
\includegraphics[width=.49\textwidth]{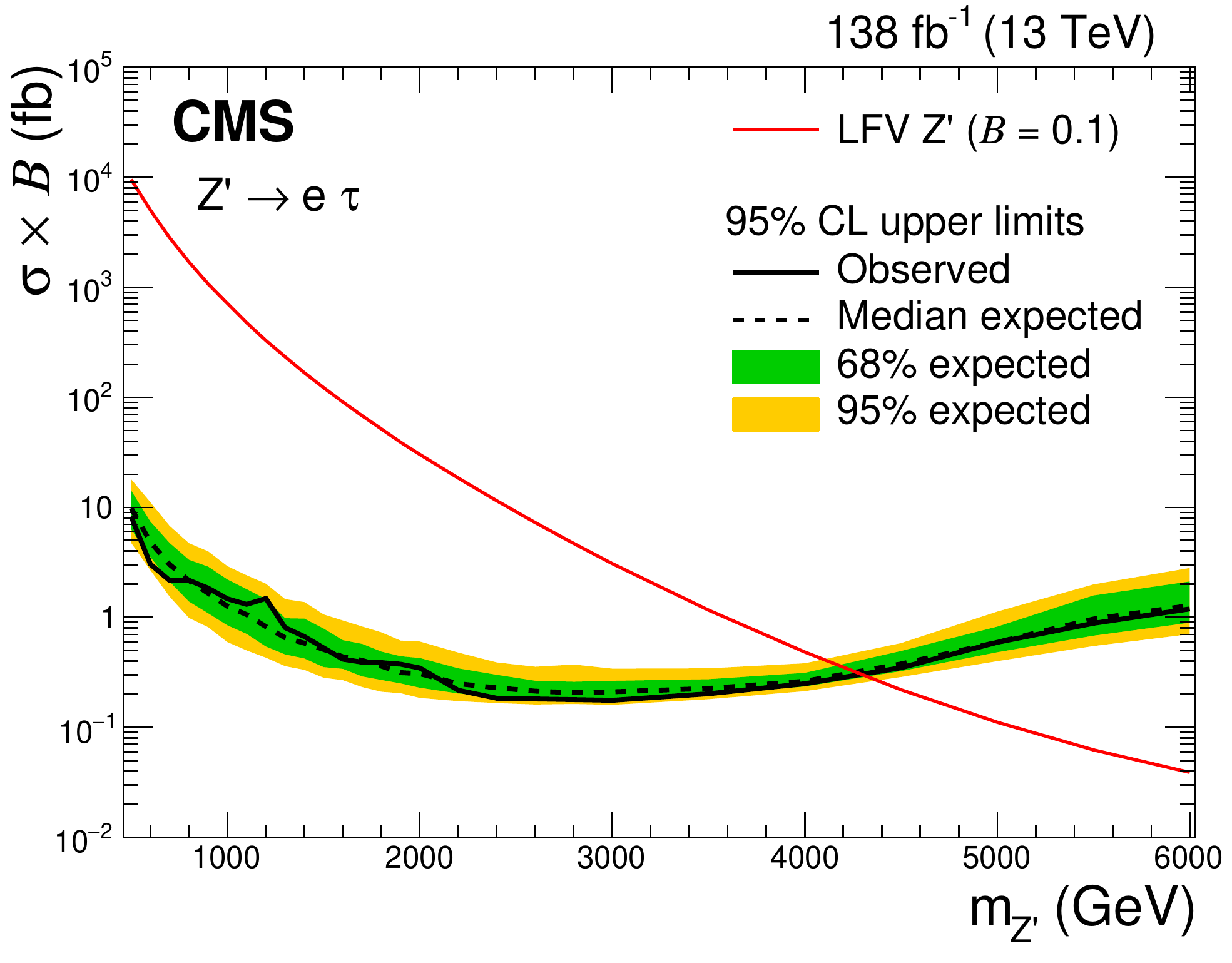}
\includegraphics[width=.49\textwidth]{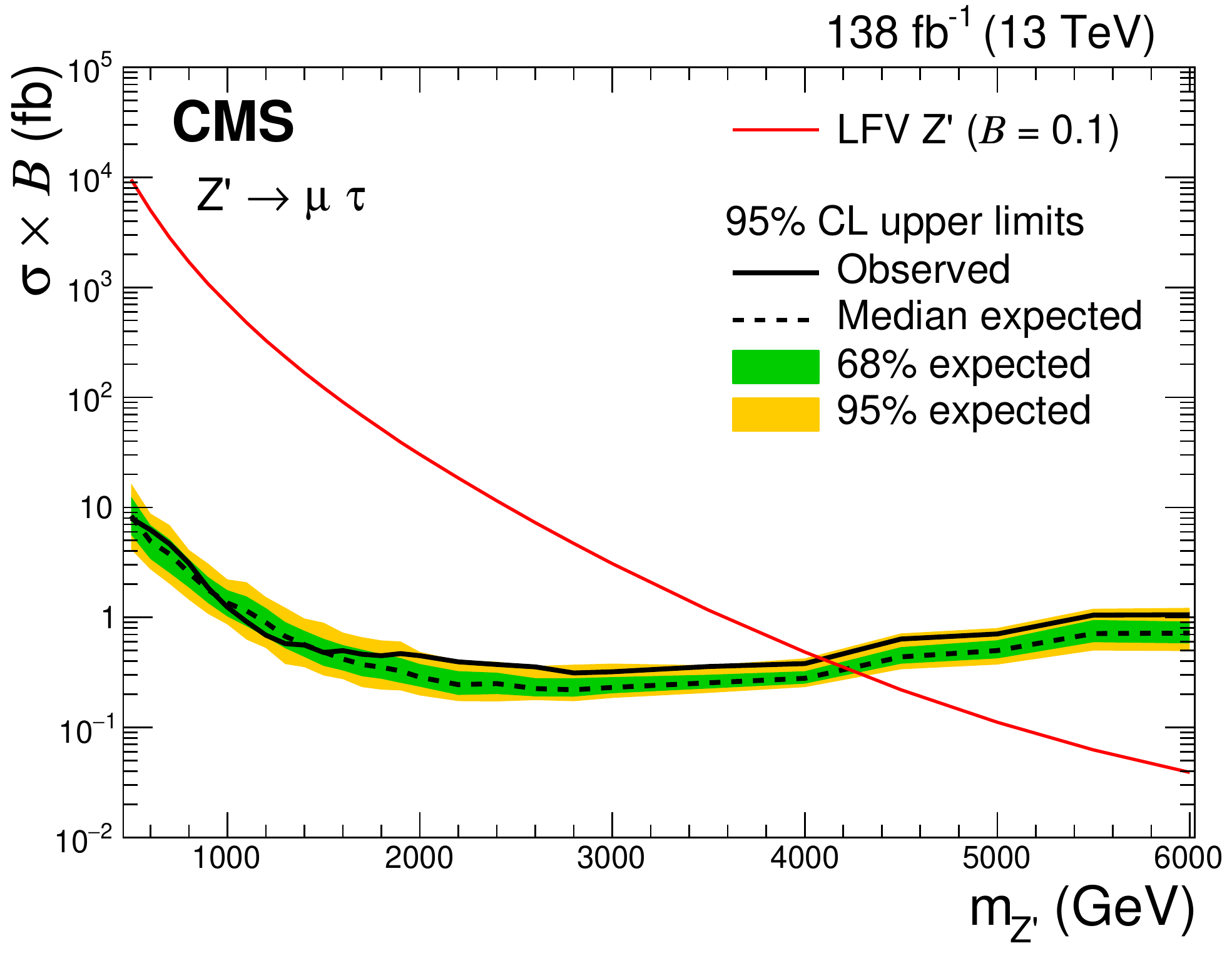}
\caption{
Expected (black dashed line) 
and observed (black solid line) 
95\% \CL  upper limits on the product of the cross section and the branching fraction
for a \PZpr boson with LFV decays, in the $\Pe\PGm$ (upper), $\Pe\PGt$ (lower left), and $\PGm\PGt$ (lower right) channels.
The shaded bands represent 
68\% and 95\% uncertainties in the expected limits. The red solid lines show the predicted product of the 
cross section and the branching fraction as a function of the 
\PZpr mass assuming $\mathcal{B}=0.1$ .
}
\label{Fig:lim_zprime}
\end{figure}

\begin{figure}[p]
\centering
\includegraphics[width=.49\textwidth]{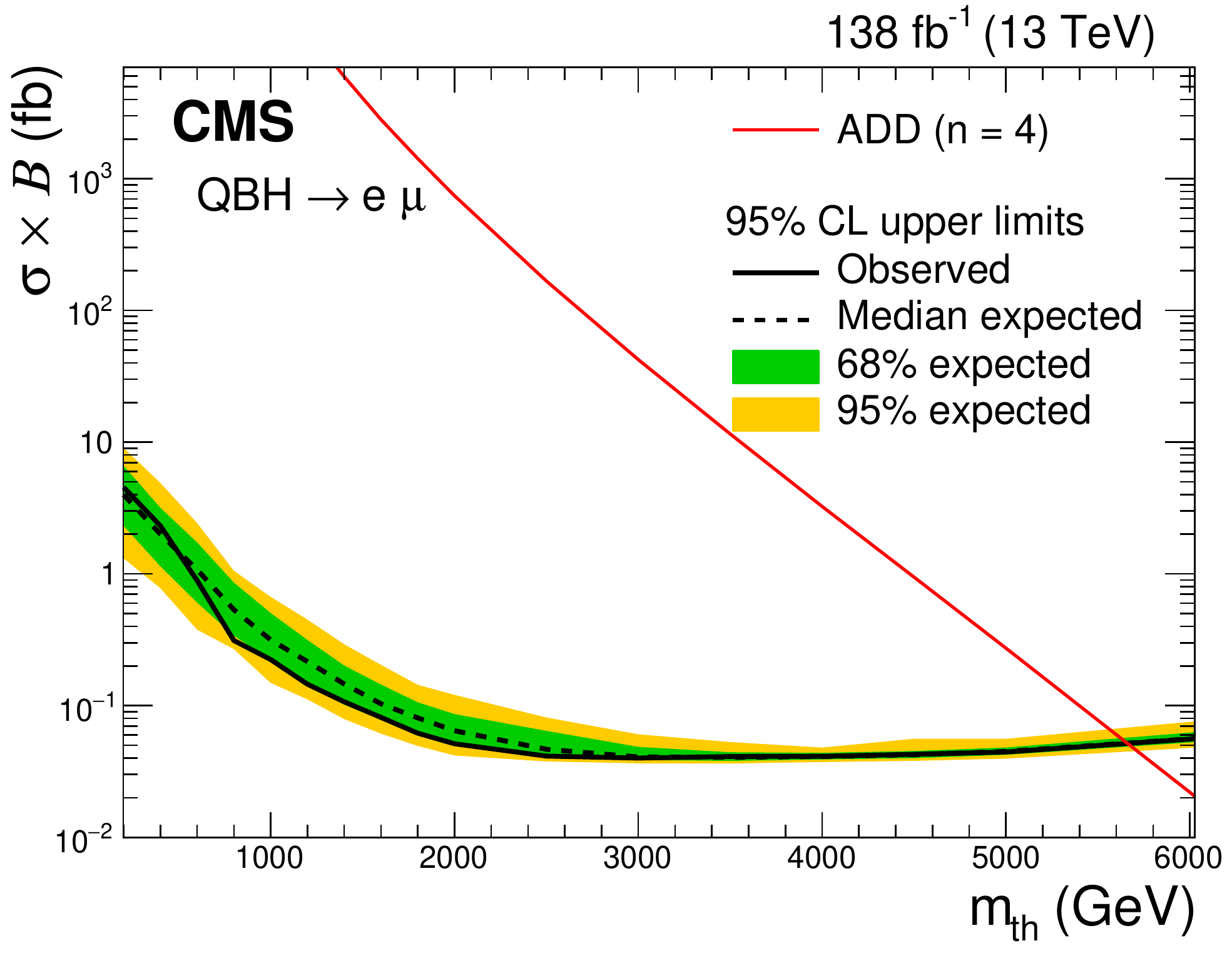} \\
\includegraphics[width=.49\textwidth]{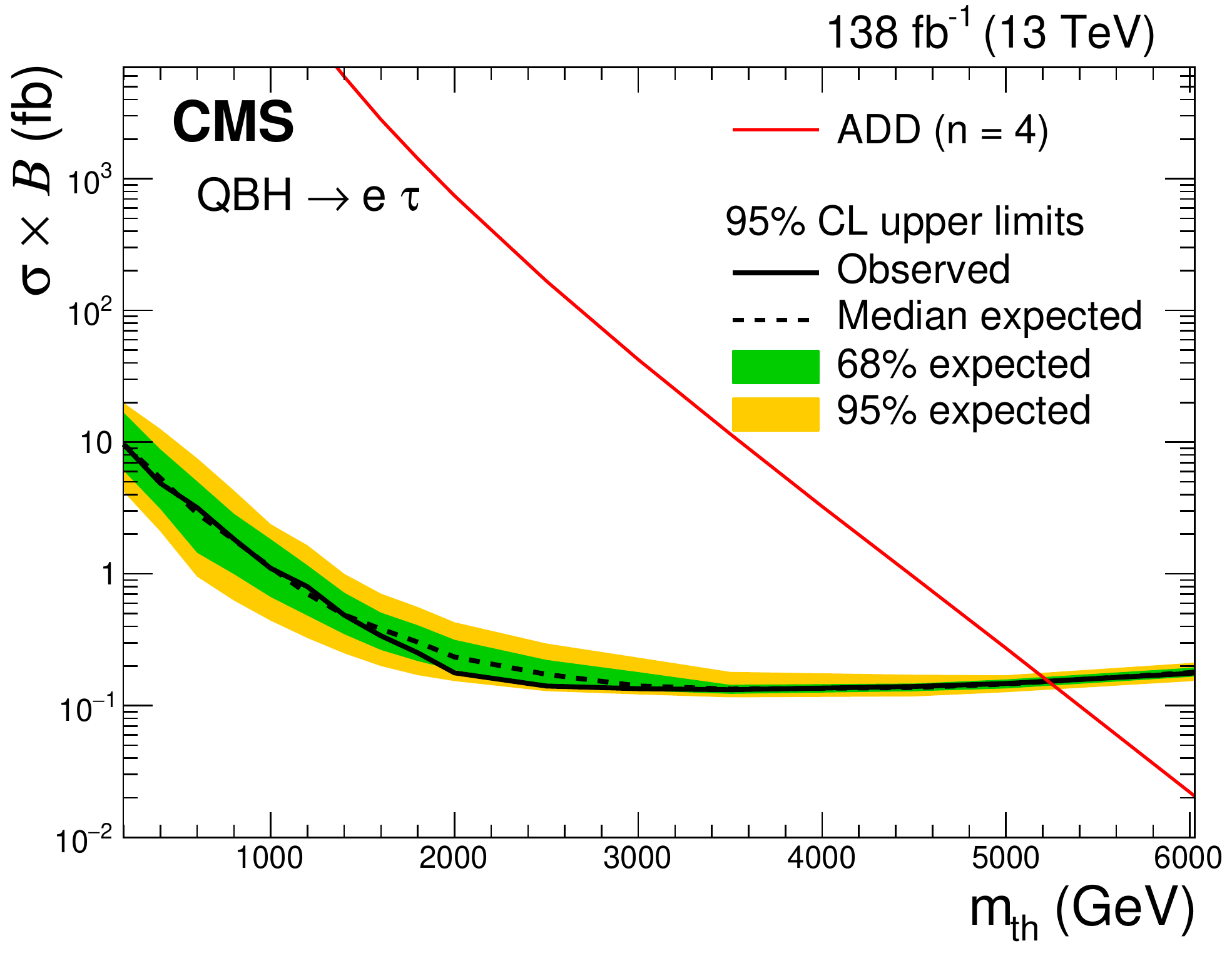}
\includegraphics[width=.49\textwidth]{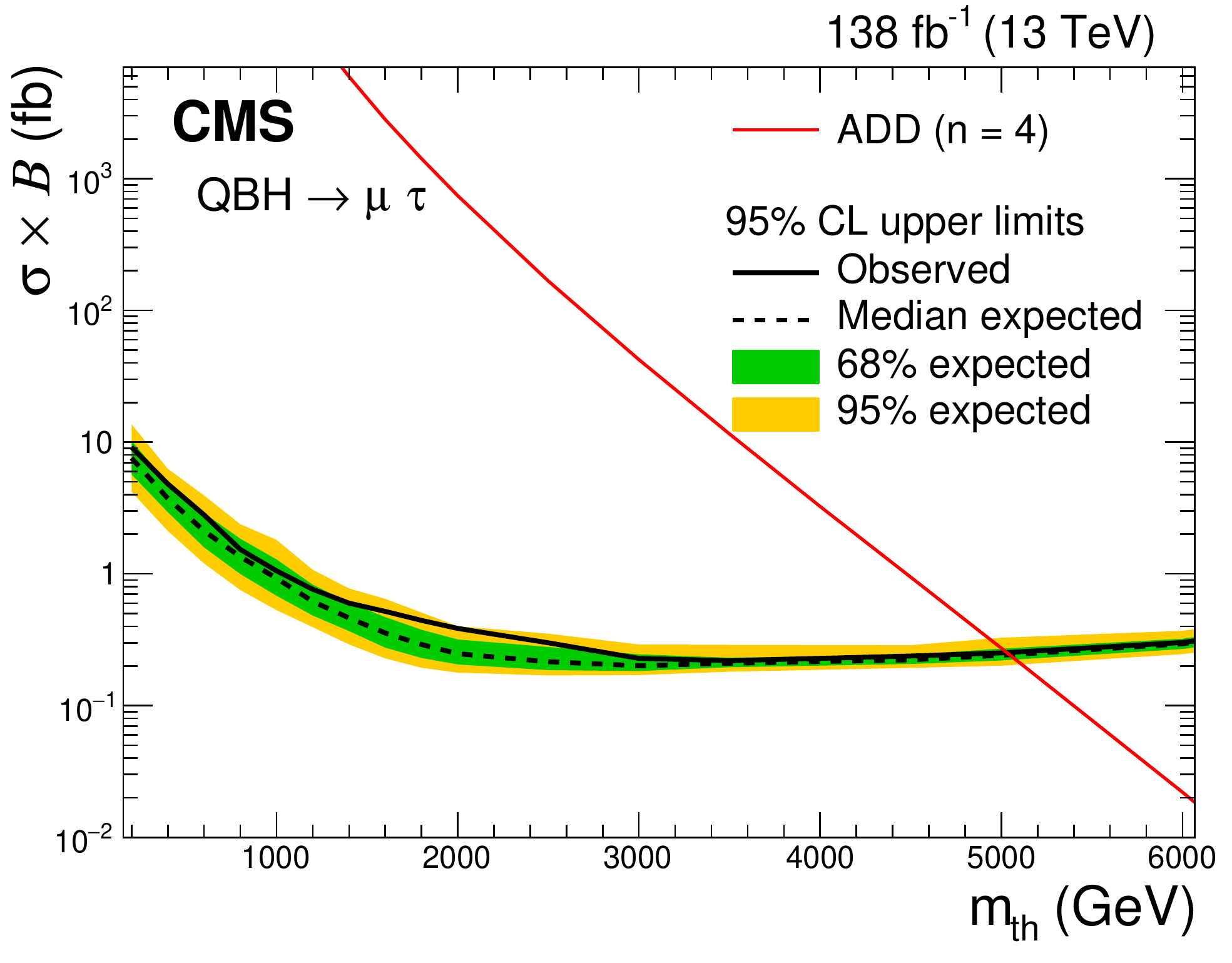}
\caption{
Expected (black dashed line) 
and observed (black solid line) 
95\% \CL upper limits on the product of the cross section and the branching fraction
for quantum black hole production in an ADD model with $n=4$ extra dimensions, in the $\Pe\PGm$ (upper), $\Pe\PGt$ (lower left), and $\PGm\PGt$ (lower right) channels.
The shaded bands represent 
68\% and 95\% uncertainties in the expected limits. 
The red solid lines show the predicted product of 
the cross section and the branching fraction as a function of the
QBH threshold mass.
}
\label{Fig:lim_qbh}
\end{figure}

\begin{figure}[p]
\centering
\includegraphics[width=.49\textwidth]{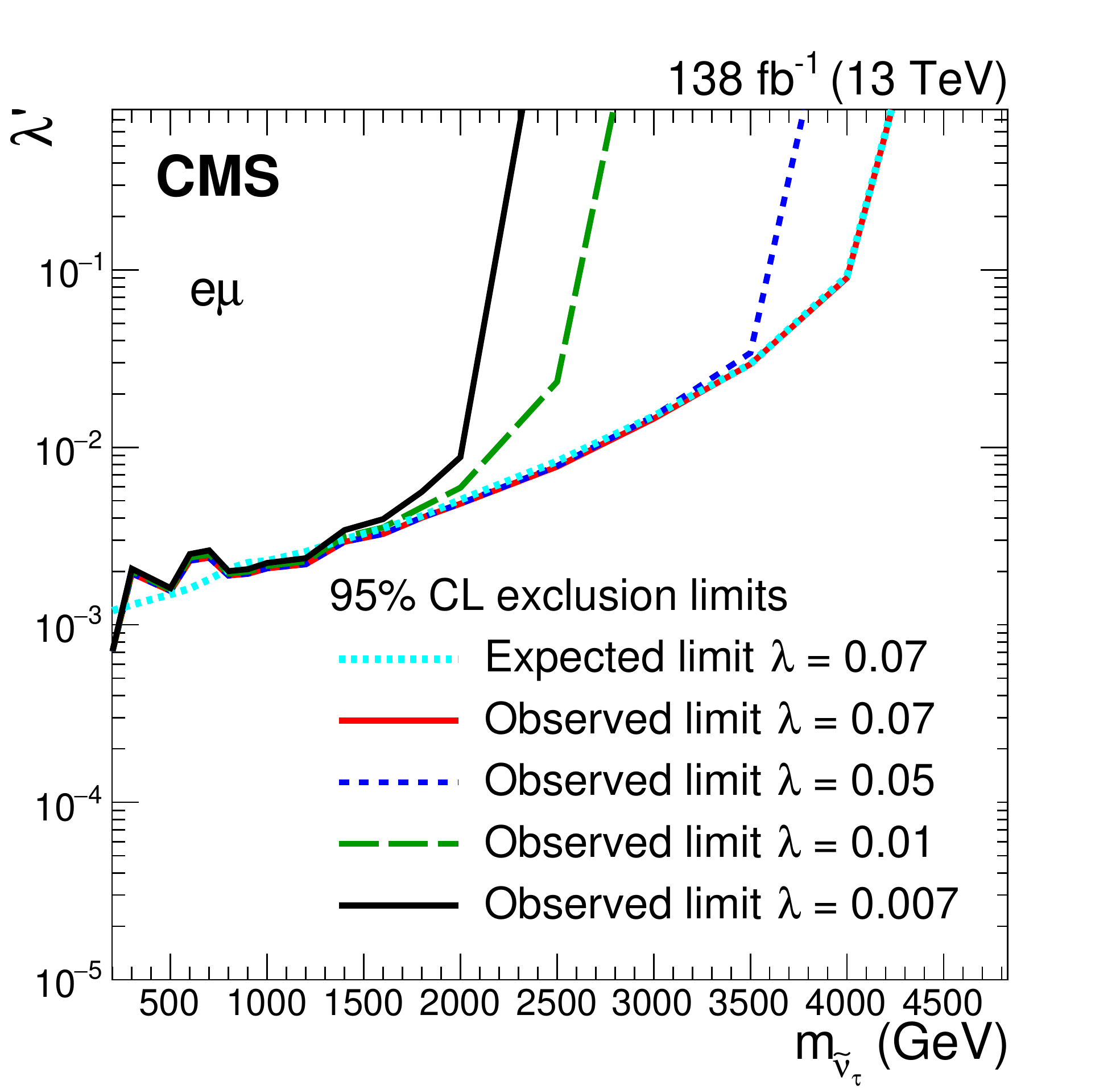} \\
\includegraphics[width=.49\textwidth]{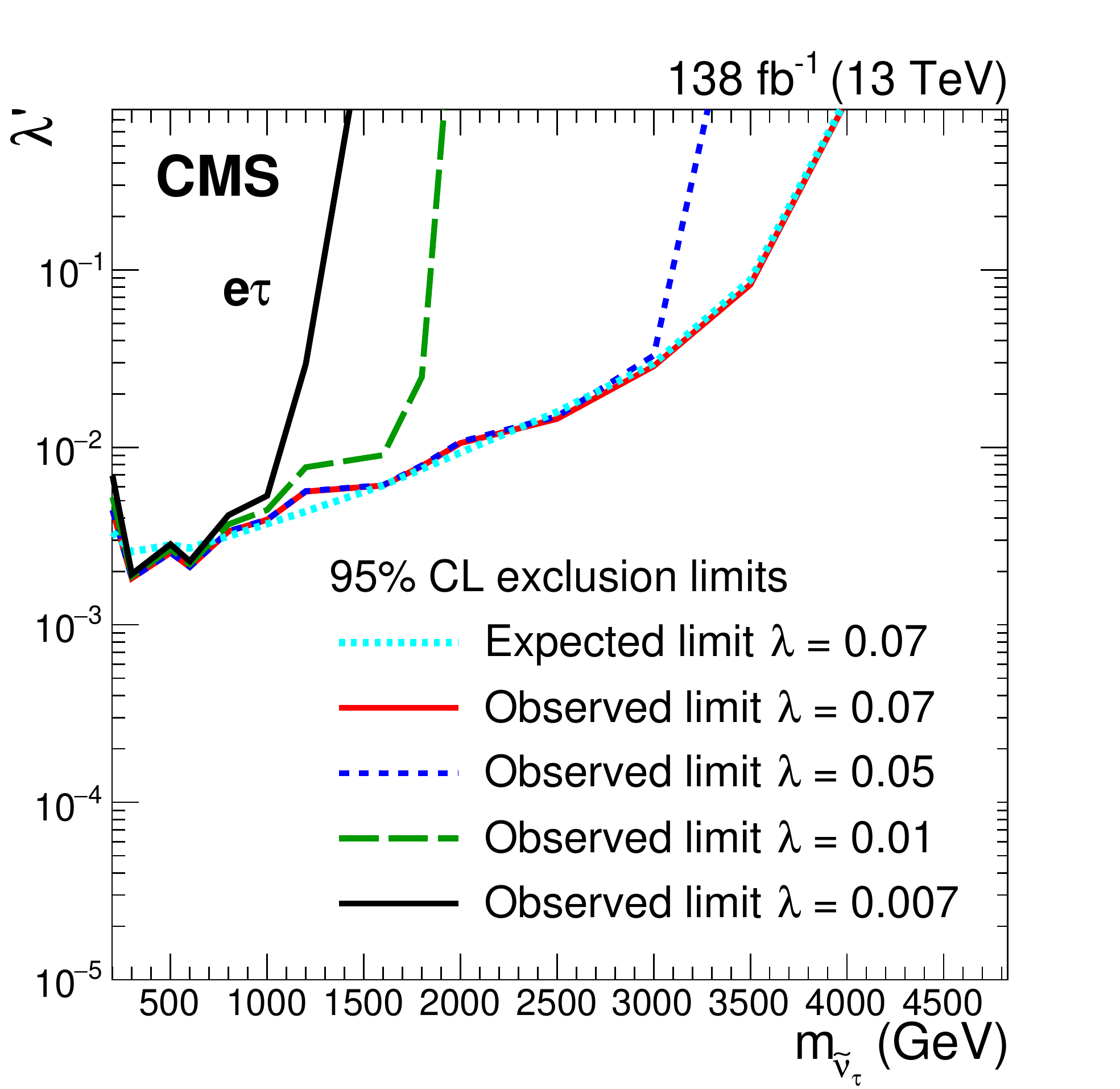}
\includegraphics[width=.49\textwidth]{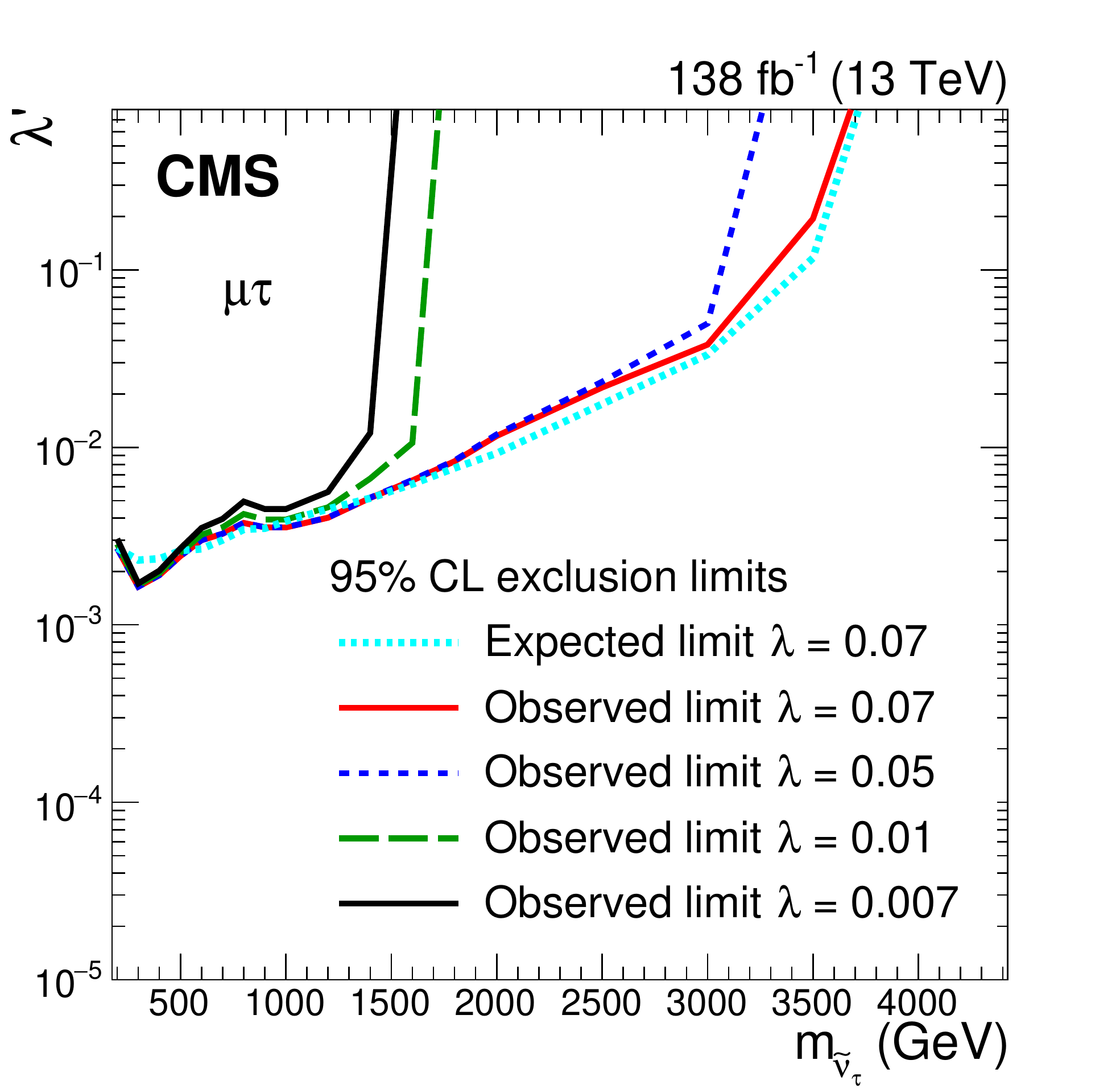}
\caption{
Exclusion limits at 95\% \CL on the RPV SUSY model in the plane of $\PGt$ sneutrino mass and $\lambda'$ coupling, for four values of $\lambda$ couplings. The regions to the left of and above the curves are
excluded. The upper plot corresponds to the $\Pe\PGm$ channel, while the lower left and right plots show the $\Pe\PGt$ and $\PGm\PGt$ channels, respectively. 
The lack of a smooth behavior of the exclusion limits for high $\lambda'$ values and at high masses where there are no events is an artifact caused by the limited number of discrete mass values of the generated signal samples, in this region.
}
\label{Fig:limit_scan}
\end{figure}

Model-specific limits for the product of the cross section 
and the branching fraction were obtained for the 
RPV SUSY, \PZpr, and QBH signals and are shown in Figs.~\ref{Fig:lim_rpv}, \ref{Fig:lim_zprime}, and \ref{Fig:lim_qbh}, respectively.
These are full Run 2 results based on 2016--2018 data sets.
The observed and expected lower mass limits obtained in all three channels are summarized in Table~\ref{tab:limit_summary}.
A $\PSgn_\PGt$ in RPV SUSY is excluded up to a mass of 4.2\TeV in the $\Pe\PGm$ channel, up to 3.7\TeV in the $\Pe\PGt$ channel, and up to 3.6\TeV for the $\PGm\PGt$ channel, in each case for the coupling hypothesis $\lambda = \lambda' = 0.1$.
If $\lambda = \lambda' = 0.01$ is assumed, the observed limits drop to 2.2\TeV in the $\Pe\PGm$ channel, 1.6\TeV in the $\Pe\PGt$ channel, and 1.6\TeV in the $\PGm\PGt$ channel. These two sets of couplings are used for the results shown in 
Figs.~\ref{Fig:lim_rpv}, \ref{Fig:lim_zprime}, and \ref{Fig:lim_qbh}, as they were for the results shown in Refs.~\cite{Sirunyan:2018zhy, Khachatryan:2016ovq}. To provide results for a wider range of couplings, in Fig.~\ref{Fig:limit_scan} we show the exclusion limits as functions of $\lambda'$, for five values of $\lambda$. 

A \PZpr ($\mathcal{B}=0.1$)  boson with LFV couplings is excluded up to a mass of 5.0\TeV in the $\Pe\PGm$ channel, up to 4.3\TeV in the $\Pe\PGt$ channel, and up to 4.1\TeV in the $\PGm\PGt$ channel. 
With increasing \PZpr boson mass,
the phase space for the on-shell \PZpr production in $\Pp\Pp$ collisions at 13\TeV decreases because of
decreasing parton-parton luminosity, leading to an increasing fraction of off-shell production at lower masses. This effect leads to weaker \PZpr boson cross section limits at higher mass
values, as shown in Fig.~\ref{Fig:lim_zprime}. 
Quantum black holes derived from an ADD model with $n=4$ extra dimensions are excluded up to the threshold mass of 5.6\TeV in the $\Pe\PGm$ channel, 5.2\TeV in the $\Pe\PGt$ channel, and 5.0\TeV in the $\PGm\PGt$ channel. 
In all channels, the results of this search significantly improve on the previous mass limits obtained by the ATLAS and CMS Collaborations.

\begin{table}[th!]
\centering
\topcaption{The observed and expected (in parentheses) 95\% \CL lower mass limits on the RPV SUSY, \PZpr, and QBH signals for the $\Pe\PGm$, $\Pe\PGt$, and $\PGm\PGt$ channels.}
\begin{tabular}{ccccc}
Channel       & \multicolumn{2}{c}{RPV SUSY $\PSgn_\PGt$ (TeV)} & LFV \PZpr (TeV) & QBH $m_{\text{th}}$ (TeV)  \\ 
              & ~~~~~$\lambda = \lambda' = 0.01$ & $\lambda = \lambda' = 0.1$~~~~~ & $\mathcal{B}=0.1$ & $n=4$\\  \hline
$\Pe\PGm$     & 2.2 (2.2)     & 4.2 (4.2)   & 5.0 (4.9)    & 5.6 (5.6)   \\
$\Pe\PGt$     & 1.6 (1.6)     & 3.7 (3.7)   & 4.3 (4.3)    & 5.2 (5.2)   \\
$\PGm\PGt$    & 1.6 (1.6)     & 3.6 (3.7)   & 4.1 (4.2)    & 5.0 (5.0)
\end{tabular}
\label{tab:limit_summary}
\end{table}

In the narrow-width approximation, the value of $\sigma \mathcal{B}$ scales with the RPV couplings, in all three channels. For example, 
in the $\Pe\PGm$ channel, the following approximation holds:
\begin{linenomath}
\begin{equation}
\sigma \mathcal{B} \approx (\lambda'_{311})^2[(\lambda_{132})^2 + (\lambda_{231})^2] / \{3 (\lambda'_{311})^2 + [ (\lambda_{132})^2 + (\lambda_{231})^2 ]  \},
\end{equation}
\end{linenomath}
Using the narrow-width approximation formula for the RPV signal cross section, the cross section
limit is translated into exclusion bounds in the plane of mass and coupling of the parameter space of the RPV SUSY
model for fixed values of the $\lambda$ couplings responsible for the decay of the $\PSgn_\PGt$. 
Limit contours in the plane of mass and coupling for
several fixed values of the coupling are shown in Fig.~\ref{Fig:limit_scan}.

\begin{figure}[p]
\centering
\includegraphics[width=.49\textwidth]{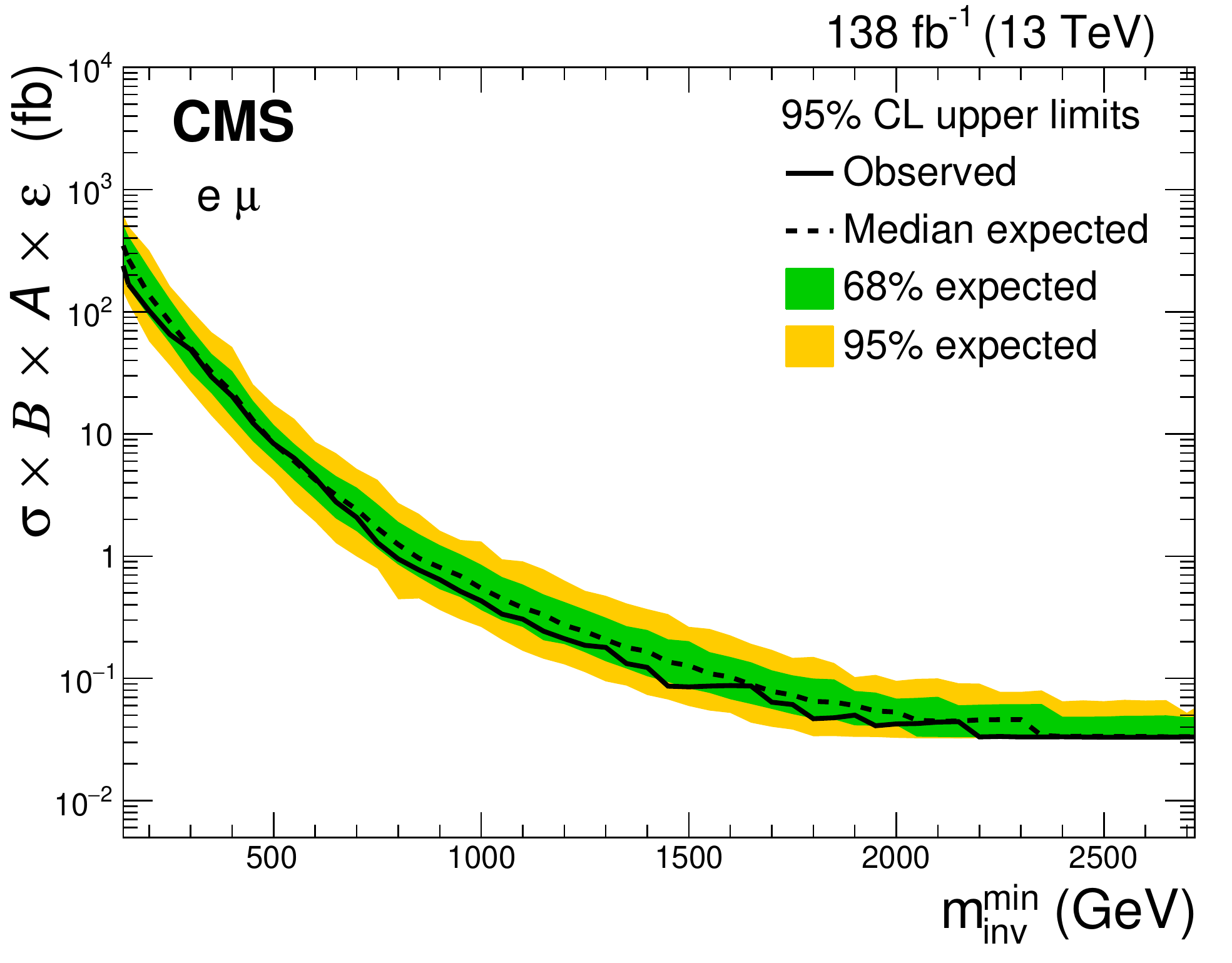} \\
\includegraphics[width=.49\textwidth]{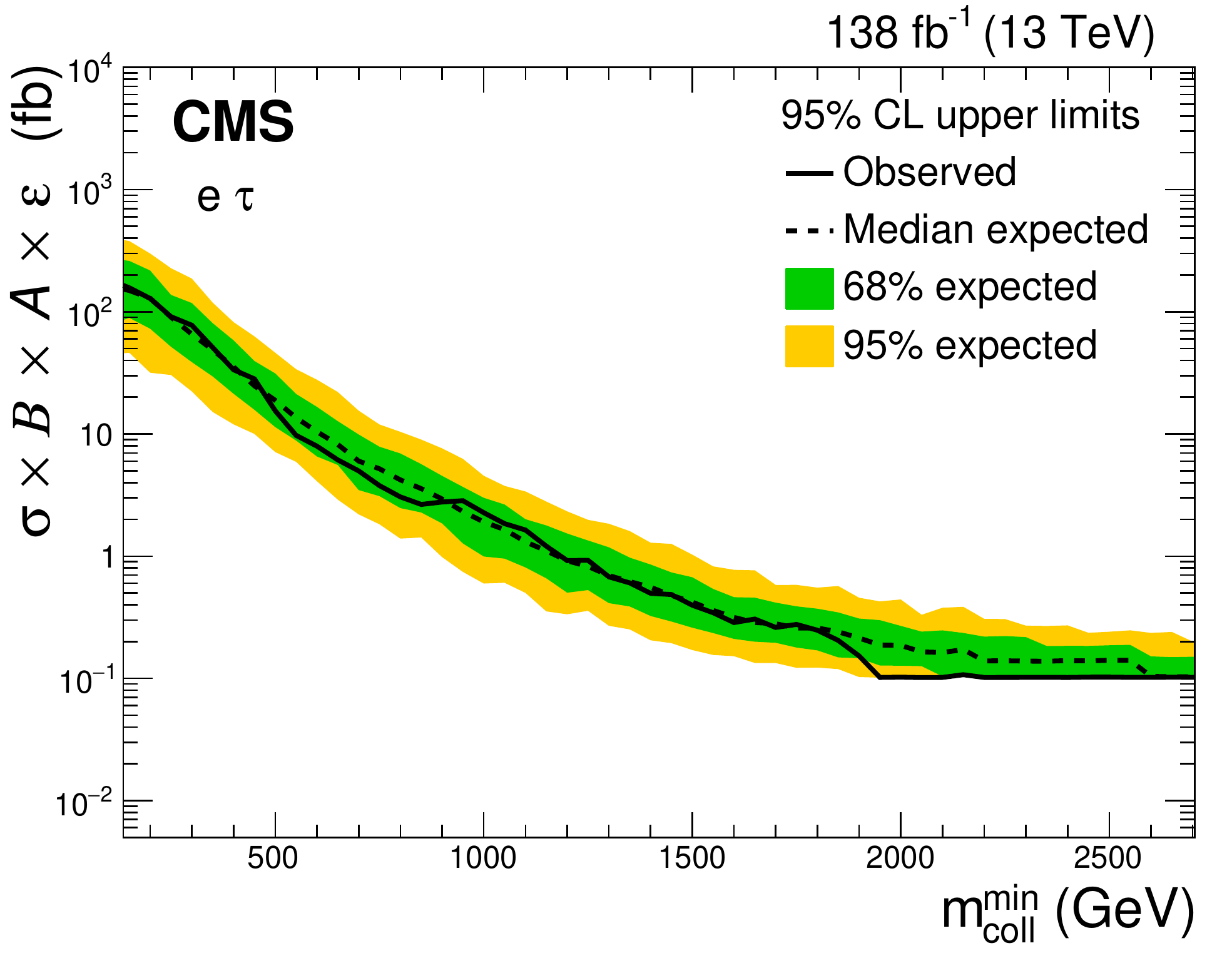}
\includegraphics[width=.49\textwidth]{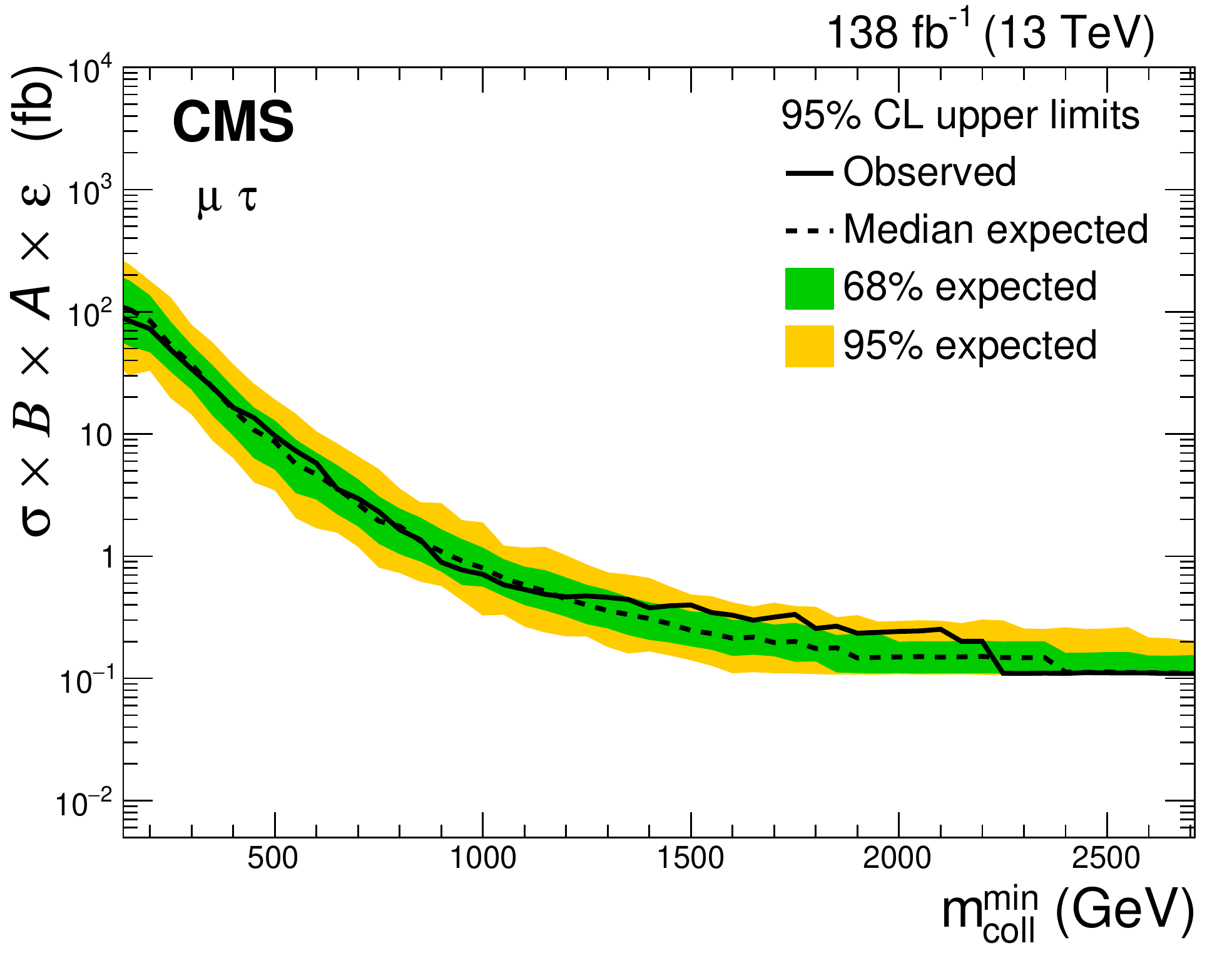}
\caption{
Model-independent upper limits at 95\% \CL on the product of the cross section, the branching fraction, acceptance, and efficiency 
are shown. Observed (expected) limits are shown in black solid (dashed) lines for the $\Pe\PGm$ (upper), $\Pe\PGt$ (lower left), and $\PGm\PGt$ (lower right) channels.
The shaded bands represent 
68\% and 95\% uncertainties in the expected limits.
}
\label{Fig:lim_mi}
\end{figure}

Model-independent cross section limits are also obtained. The model-specific shape analysis assumes a certain signal shape in invariant ($\Pe\PGm$ channel) or collinear mass ($\PGt$ channels).
However, alternative new physics processes yielding the LFV final states could cause an excess of a different shape. 
A model-independent cross section limit is determined using a single bin 
with a lower threshold on invariant (collinear) mass, and no upper threshold.
No assumptions on the shape of the signal distribution are made other than that of a flat product of acceptance times efficiency, $\text{A}\varepsilon$, 
as a function of the mass. The excluded cross section model-independent limit is shown in Fig.~\ref{Fig:lim_mi}. In order to determine the limit for a specific model from the model-independent limit described here, the model-dependent part 
of the efficiency and acceptance needs to be applied. The experimental
efficiencies for the signal are already taken into account.

A factor $f_m$ that reflects the effect of the threshold mass $m^{\text{min}}$ on the signal is determined by counting the events with masses $m>m^{\text{min}}$ and dividing 
the result by the number of MC-generated events. The reconstruction efficiency is nearly constant over the entire mass range probed here, therefore 
$f_m$ can be evaluated at the generator level. A limit on the product of the cross section and
branching fraction $(\sigma \mathcal{B} \text{A} \varepsilon)_{\text{excl}}$ can be obtained by dividing the excluded cross section of the model-independent limit  
$(\sigma \mathcal{B} \text{A} \varepsilon)_{\text{MI}}$ given in Fig.~\ref{Fig:lim_mi} by the calculated fraction $f_m$($m^{\text{min}}$):
\begin{linenomath}
\begin{equation}
(\sigma \mathcal{B} \text{A} \varepsilon)_{\text{excl}}(\text{total}) = \frac{ (\sigma \mathcal{B} \text{A} \varepsilon)_{\text{MI}} (m^{\text{min}}) }{f_m(m^{\text{min}})}.
\end{equation}
\end{linenomath}
Here, $\mathcal{B}$ is the branching fraction of the new particle decaying to the relevant LFV final state. Models with a theoretical cross section 
$(\sigma \mathcal{B})_{\text{theo}}$ larger than $(\sigma \mathcal{B})_{\text{excl}}$ can be excluded.
The fraction of events $f_m(m^{\text{min}})$ must be determined for the particular model under consideration. 

\clearpage

\section{Summary}
\label{sec_summary}
A search has been conducted for heavy particles 
that undergo lepton flavor violating decays into $\Pe\PGm$, $\Pe\PGt$, and $\PGm\PGt$ final states. The search is based on proton-proton collision data
at $\sqrt{s}=13\TeV$ recorded during 2016--2018 in the CMS detector at the CERN LHC, corresponding to an integrated luminosity 
of 138\fbinv. 
The data are consistent with expectations from the standard model. Lower limits at 95\% confidence level are set on 
the mass of supersymmetric $\PGt$ sneutrinos 
at 4.2\TeV in $\Pe\PGm$, 3.7\TeV in $\Pe\PGt$, and 3.6\TeV in $\PGm\PGt$ channels.
A \PZpr vector boson 
with lepton flavor violating couplings is excluded for masses below 5.0, 4.3, and 4.1\TeV in the $\Pe\PGm$, $\Pe\PGt$, and $\PGm\PGt$ channels, respectively, assuming
a branching fraction of 10\%. 
In the context of the Arkani-Hamed--Dimopoulos--Dvali model with four extra dimensions, values of the threshold mass for quantum black hole production 
less than 5.6, 5.2, and 5.0\TeV are excluded in the $\Pe\PGm$, $\Pe\PGt$, and $\PGm\PGt$ channels, 
respectively. 
In addition, model-independent limits are provided allowing the results to be interpreted in other models with the same final states and similar kinematic distributions. Limits in the $\Pe\PGt$ and $\PGm\PGt$ final states, as well as model-independent limits, are reported for 
the first time in the context of a high-mass lepton flavor violation search. 
These are the first results of a high-mass lepton flavor violation search using the full Run 2 data set, and they 
are currently the most stringent limits from any collider experiment. 

\begin{acknowledgments}
We congratulate our colleagues in the CERN accelerator departments for the excellent performance of the LHC and thank the technical and administrative staffs at CERN and at other CMS institutes for their contributions to the success of the CMS effort. In addition, we gratefully acknowledge the computing centers and personnel of the Worldwide LHC Computing Grid and other centers for delivering so effectively the computing infrastructure essential to our analyses. Finally, we acknowledge the enduring support for the construction and operation of the LHC, the CMS detector, and the supporting computing infrastructure provided by the following funding agencies: BMBWF and FWF (Austria); FNRS and FWO (Belgium); CNPq, CAPES, FAPERJ, FAPERGS, and FAPESP (Brazil); MES and BNSF (Bulgaria); CERN; CAS, MoST, and NSFC (China); MINCIENCIAS (Colombia); MSES and CSF (Croatia); RIF (Cyprus); SENESCYT (Ecuador); MoER, ERC PUT and ERDF (Estonia); Academy of Finland, MEC, and HIP (Finland); CEA and CNRS/IN2P3 (France); BMBF, DFG, and HGF (Germany); GSRI (Greece); NKFIH (Hungary); DAE and DST (India); IPM (Iran); SFI (Ireland); INFN (Italy); MSIP and NRF (Republic of Korea); MES (Latvia); LAS (Lithuania); MOE and UM (Malaysia); BUAP, CINVESTAV, CONACYT, LNS, SEP, and UASLP-FAI (Mexico); MOS (Montenegro); MBIE (New Zealand); PAEC (Pakistan); MES and NSC (Poland); FCT (Portugal); MESTD (Serbia); MCIN/AEI and PCTI (Spain); MOSTR (Sri Lanka); Swiss Funding Agencies (Switzerland); MST (Taipei); MHESI and NSTDA (Thailand); TUBITAK and TENMAK (Turkey); NASU (Ukraine); STFC (United Kingdom); DOE and NSF (USA).

\hyphenation{Rachada-pisek} Individuals have received support from the Marie-Curie program and the European Research Council and Horizon 2020 Grant, contract Nos.\ 675440, 724704, 752730, 758316, 765710, 824093, 884104, and COST Action CA16108 (European Union); the Leventis Foundation; the Alfred P.\ Sloan Foundation; the Alexander von Humboldt Foundation; the Belgian Federal Science Policy Office; the Fonds pour la Formation \`a la Recherche dans l'Industrie et dans l'Agriculture (FRIA-Belgium); the Agentschap voor Innovatie door Wetenschap en Technologie (IWT-Belgium); the F.R.S.-FNRS and FWO (Belgium) under the ``Excellence of Science -- EOS" -- be.h project n.\ 30820817; the Beijing Municipal Science \& Technology Commission, No. Z191100007219010; the Ministry of Education, Youth and Sports (MEYS) of the Czech Republic; the Hellenic Foundation for Research and Innovation (HFRI), Project Number 2288 (Greece); the Deutsche Forschungsgemeinschaft (DFG), under Germany's Excellence Strategy -- EXC 2121 ``Quantum Universe" -- 390833306, and under project number 400140256 - GRK2497; the Hungarian Academy of Sciences, the New National Excellence Program - \'UNKP, the NKFIH research grants K 124845, K 124850, K 128713, K 128786, K 129058, K 131991, K 133046, K 138136, K 143460, K 143477, 2020-2.2.1-ED-2021-00181, and TKP2021-NKTA-64 (Hungary); the Council of Science and Industrial Research, India; the Latvian Council of Science; the Ministry of Education and Science, project no. 2022/WK/14, and the National Science Center, contracts Opus 2021/41/B/ST2/01369 and 2021/43/B/ST2/01552 (Poland); the Funda\c{c}\~ao para a Ci\^encia e a Tecnologia, grant CEECIND/01334/2018 (Portugal); the National Priorities Research Program by Qatar National Research Fund; MCIN/AEI/10.13039/501100011033, ERDF ``a way of making Europe", and the Programa Estatal de Fomento de la Investigaci{\'o}n Cient{\'i}fica y T{\'e}cnica de Excelencia Mar\'{\i}a de Maeztu, grant MDM-2017-0765 and Programa Severo Ochoa del Principado de Asturias (Spain); the Chulalongkorn Academic into Its 2nd Century Project Advancement Project, and the National Science, Research and Innovation Fund via the Program Management Unit for Human Resources \& Institutional Development, Research and Innovation, grant B05F650021 (Thailand); the Kavli Foundation; the Nvidia Corporation; the SuperMicro Corporation; the Welch Foundation, contract C-1845; and the Weston Havens Foundation (USA).    
\end{acknowledgments}

\bibliography{auto_generated}  

\providecommand{\href}[2]{#2}\begingroup\raggedright\begin{thebibliography}{10}%
\makeatletter
\providecommand{\hrefCMSnoop }[0]{\@secondoftwo}%
\makeatother
\providecommand{\doi}{\texttt{doi:}\begingroup \urlstyle{tt}\Url}

\bibitem{Farrar:1978xj}
\hrefCMSnoop {}{G.~R. Farrar and P.~Fayet, ``Phenomenology of the production,
  decay, and detection of new hadronic states associated with supersymmetry'',}
  \textit{ Phys. Lett. B} \textbf{ 76} (1978) 575,
\href{http://dx.doi.org/10.1016/0370-2693(78)90858-4}{\doi{10.1016/0370-2693(78)90858-4}}.
%%CITATION = PHLTA,76B,575;%%.

\bibitem{deCampos:2008av}
F.~de~Campos\hrefCMSnoop {}{ {et~al.}, ``{CERN} {LHC} signals for neutrino mass
  model in bilinear {R}-parity violating {mAMSB}'',} \textit{ Phys. Rev. D}
  \textbf{ 77} (2008) 115025,
  \href{http://dx.doi.org/10.1103/PhysRevD.77.115025}{\doi{10.1103/PhysRevD.77.115025}},
\href{http://www.arXiv.org/abs/0803.4405}{\texttt{arXiv:0803.4405}}.
%%CITATION = ARXIV:0803.4405;%%.

\bibitem{Barbier:2004ez}
R.~Barbier\hrefCMSnoop {}{ {et~al.}, ``{R}-parity violating supersymmetry'',}
  \textit{ Phys. Rept.} \textbf{ 420} (2005) 1,
  \href{http://dx.doi.org/10.1016/j.physrep.2005.08.006}{\doi{10.1016/j.physrep.2005.08.006}},
\href{http://www.arXiv.org/abs/hep-ph/0406039}{\texttt{arXiv:hep-ph/0406039}}.
%%CITATION = HEP-PH/0406039;%%.

\bibitem{Langacker:2008yv}
\hrefCMSnoop {}{P.~Langacker, ``The physics of heavy {$\PZ^\prime$} gauge
  bosons'',} \textit{ Rev. Mod. Phys.} \textbf{ 81} (2009) 1199,
  \href{http://dx.doi.org/10.1103/RevModPhys.81.1199}{\doi{10.1103/RevModPhys.81.1199}},
\href{http://www.arXiv.org/abs/0801.1345}{\texttt{arXiv:0801.1345}}.
%%CITATION = ARXIV:0801.1345;%%.

\bibitem{Calmet:2008dg}
\hrefCMSnoop {}{X.~Calmet, W.~Gong, and S.~D.~H. Hsu, ``Colorful quantum black
  holes at the {LHC}'',} \textit{ Phys. Lett. B} \textbf{ 668} (2008) 20,
  \href{http://dx.doi.org/10.1016/j.physletb.2008.08.011}{\doi{10.1016/j.physletb.2008.08.011}},
\href{http://www.arXiv.org/abs/0806.4605}{\texttt{arXiv:0806.4605}}.
%%CITATION = ARXIV:0806.4605;%%.

\bibitem{Meade:2007sz}
\hrefCMSnoop {}{P.~Meade and L.~Randall, ``Black holes and quantum gravity at
  the {LHC}'',} \textit{ JHEP} \textbf{ 05} (2008) 003,
  \href{http://dx.doi.org/10.1088/1126-6708/2008/05/003}{\doi{10.1088/1126-6708/2008/05/003}},
\href{http://www.arXiv.org/abs/0708.3017}{\texttt{arXiv:0708.3017}}.
%%CITATION = ARXIV:0708.3017;%%.

\bibitem{Dimopoulos:2001hw}
\hrefCMSnoop {}{S.~Dimopoulos and G.~Landsberg, ``Black holes at the {LHC}'',}
  \textit{ Phys. Rev. Lett.} \textbf{ 87} (2001) 161602,
  \href{http://dx.doi.org/10.1103/PhysRevLett.87.161602}{\doi{10.1103/PhysRevLett.87.161602}},
  \href{http://www.arXiv.org/abs/hep-ph/0106295}{\texttt{arXiv:hep-ph/0106295}}.

\bibitem{Giddings:2001bu}
\hrefCMSnoop {}{S.~B. Giddings and S.~D. Thomas, ``High-energy colliders as
  black hole factories: The end of short distance physics'',} \textit{ Phys.
  Rev. D} \textbf{ 65} (2002) 056010,
  \href{http://dx.doi.org/10.1103/PhysRevD.65.056010}{\doi{10.1103/PhysRevD.65.056010}},
  \href{http://www.arXiv.org/abs/hep-ph/0106219}{\texttt{arXiv:hep-ph/0106219}}.

\bibitem{Graham:2012th}
\hrefCMSnoop {}{P.~W. Graham, D.~E. Kaplan, S.~Rajendran, and P.~Saraswat,
  ``Displaced supersymmetry'',} \textit{ JHEP} \textbf{ 07} (2012) 149,
  \href{http://dx.doi.org/10.1007/JHEP07(2012)149}{\doi{10.1007/JHEP07(2012)149}},
\href{http://www.arXiv.org/abs/1204.6038}{\texttt{arXiv:1204.6038}}.
%%CITATION = ARXIV:1204.6038;%%.

\bibitem{Sirunyan:2018zhy}
\hrefCMSnoop {}{{CMS Collaboration}, ``Search for lepton-flavor violating
  decays of heavy resonances and quantum black holes to {$\Pe\Pgm$} final
  states in proton-proton collisions at {$\sqrt{s}=13\TeV$}'',} \textit{ JHEP}
  \textbf{ 04} (2018) 073,
  \href{http://dx.doi.org/10.1007/JHEP04(2018)073}{\doi{10.1007/JHEP04(2018)073}},
\href{http://www.arXiv.org/abs/1802.01122}{\texttt{arXiv:1802.01122}}.
%%CITATION = ARXIV:1802.01122;%%.

\bibitem{ATLAS:2018mrn}
\hrefCMSnoop {}{{ATLAS Collaboration}, ``Search for lepton-flavor violation in
  different-flavor, high-mass final states in pp collisions at {$\sqrt
  s=13\TeV$} with the {ATLAS} detector'',} \textit{ Phys. Rev. D} \textbf{ 98}
  (2018) 092008,
  \href{http://dx.doi.org/10.1103/PhysRevD.98.092008}{\doi{10.1103/PhysRevD.98.092008}},
  \href{http://www.arXiv.org/abs/1807.06573}{\texttt{arXiv:1807.06573}}.

\bibitem{Gingrich}
\hrefCMSnoop {}{D.~M. Gingrich, ``Quantum black holes with charge, colour, and
  spin at the {LHC}'',} \textit{ J. Phys. G} \textbf{ 37} (2010) 105008,
  \href{http://dx.doi.org/10.1088/0954-3899/37/10/105008}{\doi{10.1088/0954-3899/37/10/105008}},
\href{http://www.arXiv.org/abs/0912.0826}{\texttt{arXiv:0912.0826}}.
%%CITATION = ARXIV:0912.0826;%%.

\bibitem{Randall:1999ee}
\hrefCMSnoop {}{L.~Randall and R.~Sundrum, ``A large mass hierarchy from a
  small extra dimension'',} \textit{ Phys. Rev. Lett.} \textbf{ 83} (1999)
  3370,
  \href{http://dx.doi.org/10.1103/PhysRevLett.83.3370}{\doi{10.1103/PhysRevLett.83.3370}},
\href{http://www.arXiv.org/abs/hep-ph/9905221}{\texttt{arXiv:hep-ph/9905221}}.
%%CITATION = HEP-PH/9905221;%%.

\bibitem{ArkaniHamed:1998rs}
\hrefCMSnoop {}{N.~Arkani-Hamed, S.~Dimopoulos, and G.~R. Dvali, ``The
  hierarchy problem and new dimensions at a millimeter'',} \textit{ Phys. Lett.
  B} \textbf{ 429} (1998) 263,
  \href{http://dx.doi.org/10.1016/S0370-2693(98)00466-3}{\doi{10.1016/S0370-2693(98)00466-3}},
\href{http://www.arXiv.org/abs/hep-ph/9803315}{\texttt{arXiv:hep-ph/9803315}}.
%%CITATION = HEP-PH/9803315;%%.

\bibitem{Aaltonen:2010fv}
\hrefCMSnoop {}{{CDF} Collaboration, ``Search for {R}-parity violating decays
  of {$\Pgt$} sneutrinos to {$\Pe\Pgm$}, {$\Pgm\Pgt$}, and {$\Pe\Pgt$} pairs in
  {$\Pp\bar{\Pp}$} collisions at {$\sqrt{s} = 1.96\TeV$}'',} \textit{ Phys.
  Rev. Lett.} \textbf{ 105} (2010) 191801,
  \href{http://dx.doi.org/10.1103/PhysRevLett.105.191801}{\doi{10.1103/PhysRevLett.105.191801}},
\href{http://www.arXiv.org/abs/1004.3042}{\texttt{arXiv:1004.3042}}.
%%CITATION = ARXIV:1004.3042;%%.

\bibitem{Abazov:2010km}
\hrefCMSnoop {}{{D0} Collaboration, ``Search for sneutrino production in
  {$\Pe\Pgm$} final states in 5.3\fbinv of {$\Pp\bar{\Pp}$} collisions at
  {$\sqrt{s} = 1.96\TeV$}'',} \textit{ Phys. Rev. Lett.} \textbf{ 105} (2010)
  191802,
  \href{http://dx.doi.org/10.1103/PhysRevLett.105.191802}{\doi{10.1103/PhysRevLett.105.191802}},
\href{http://www.arXiv.org/abs/1007.4835}{\texttt{arXiv:1007.4835}}.
%%CITATION = ARXIV:1007.4835;%%.

\bibitem{Aad:2015pfa}
\hrefCMSnoop {}{{ATLAS Collaboration}, ``Search for a heavy neutral particle
  decaying to {$\Pe\Pgm$}, {$\Pe\Pgt$}, or {$\Pgm\Pgt$} in {$\Pp\Pp$}
  collisions at {$\sqrt{s}=8\TeV$} with the {ATLAS} detector'',} \textit{ Phys.
  Rev. Lett.} \textbf{ 115} (2015) 031801,
  \href{http://dx.doi.org/10.1103/PhysRevLett.115.031801}{\doi{10.1103/PhysRevLett.115.031801}},
\href{http://www.arXiv.org/abs/1503.04430}{\texttt{arXiv:1503.04430}}.
%%CITATION = ARXIV:1503.04430;%%.

\bibitem{Khachatryan:2016ovq}
\hrefCMSnoop {}{{CMS Collaboration}, ``Search for lepton flavour violating
  decays of heavy resonances and quantum black holes to an {$\Pe\Pgm$} pair in
  proton-proton collisions at {$\sqrt{s} = 8\TeV$}'',} \textit{ Eur. Phys. J.
  C} \textbf{ 76} (2016) 317,
  \href{http://dx.doi.org/10.1140/epjc/s10052-016-4149-y}{\doi{10.1140/epjc/s10052-016-4149-y}},
\href{http://www.arXiv.org/abs/1604.05239}{\texttt{arXiv:1604.05239}}.
%%CITATION = ARXIV:1604.05239;%%.

\bibitem{TheATLAScollaboration:2015ucd}
\hrefCMSnoop {}{{ATLAS Collaboration}, ``Search for new phenomena in
  different-flavour high-mass dilepton final states in pp collisions at
  {$\sqrt{s}=13\TeV$} with the {ATLAS} detector'',} \textit{ Eur. Phys. J. C}
  \textbf{ 76} (2016) 541,
  \href{http://dx.doi.org/10.1140/epjc/s10052-016-4385-1}{\doi{10.1140/epjc/s10052-016-4385-1}},
\href{http://www.arXiv.org/abs/1607.08079}{\texttt{arXiv:1607.08079}}.
%%CITATION = ARXIV:1607.08079;%%.

\bibitem{hepdata}
\href {https://doi.org/10.17182/hepdata.127302}{{CMS Collaboration}, ``Hepdata
  record for this analysis'',} Technical Report doi::10.17182/hepdata.127302,
  2022.

\bibitem{TRK-11-001}
\hrefCMSnoop {}{{CMS Collaboration}, ``Description and performance of track and
  primary-vertex reconstruction with the {CMS} tracker'',} \textit{ JINST}
  \textbf{ 9} (2014) P10009,
  \href{http://dx.doi.org/10.1088/1748-0221/9/10/P10009}{\doi{10.1088/1748-0221/9/10/P10009}},
\href{http://www.arXiv.org/abs/1405.6569}{\texttt{arXiv:1405.6569}}.
%%CITATION = ARXIV:1405.6569;%%.

\bibitem{CMS:2020cmk}
\hrefCMSnoop {}{{CMS Collaboration}, ``Performance of the {CMS Level-1} trigger
  in proton-proton collisions at {$\sqrt{s} = 13\TeV$}'',} \textit{ JINST}
  \textbf{ 15} (2020) P10017,
  \href{http://dx.doi.org/10.1088/1748-0221/15/10/P10017}{\doi{10.1088/1748-0221/15/10/P10017}},
  \href{http://www.arXiv.org/abs/2006.10165}{\texttt{arXiv:2006.10165}}.

\bibitem{CMS:2016ngn}
\hrefCMSnoop {}{{CMS Collaboration}, ``The {CMS} trigger system'',} \textit{
  JINST} \textbf{ 12} (2017) P01020,
  \href{http://dx.doi.org/10.1088/1748-0221/12/01/P01020}{\doi{10.1088/1748-0221/12/01/P01020}},
\href{http://www.arXiv.org/abs/1609.02366}{\texttt{arXiv:1609.02366}}.
%%CITATION = ARXIV:1609.02366;%%.

\bibitem{muon_paper}
\hrefCMSnoop {}{{CMS Collaboration}, ``Performance of the {CMS} muon detector
  and muon reconstruction with proton-proton collisions at {$\sqrt{s} =
  13\TeV$}'',} \textit{ JINST} \textbf{ 13} (2018) P06015,
  \href{http://dx.doi.org/10.1088/1748-0221/13/06/P06015}{\doi{10.1088/1748-0221/13/06/P06015}},
\href{http://www.arXiv.org/abs/1804.04528}{\texttt{arXiv:1804.04528}}.
%%CITATION = ARXIV:1804.04528;%%.

\bibitem{CMS:2020uim}
\hrefCMSnoop {}{{CMS Collaboration}, ``Electron and photon reconstruction and
  identification with the {CMS} experiment at the {CERN LHC}'',} \textit{
  JINST} \textbf{ 16} (2021) P05014,
  \href{http://dx.doi.org/10.1088/1748-0221/16/05/P05014}{\doi{10.1088/1748-0221/16/05/P05014}},
  \href{http://www.arXiv.org/abs/2012.06888}{\texttt{arXiv:2012.06888}}.

\bibitem{Khachatryan:2015hwa}
\hrefCMSnoop {}{{CMS Collaboration}, ``Performance of electron reconstruction
  and selection with the {CMS} detector in proton-proton collisions at
  {$\sqrt{s} = 8\TeV$}'',} \textit{ JINST} \textbf{ 10} (2015) P06005,
  \href{http://dx.doi.org/10.1088/1748-0221/10/06/P06005}{\doi{10.1088/1748-0221/10/06/P06005}},
\href{http://www.arXiv.org/abs/1502.02701}{\texttt{arXiv:1502.02701}}.
%%CITATION = ARXIV:1502.02701;%%.

\bibitem{Chatrchyan:2008zzk}
\hrefCMSnoop {}{{CMS Collaboration}, ``The {CMS} experiment at the {CERN}
  {LHC}'',} \textit{ JINST} \textbf{ 3} (2008) S08004,
  \href{http://dx.doi.org/10.1088/1748-0221/3/08/S08004}{\doi{10.1088/1748-0221/3/08/S08004}}.

\bibitem{Belyaev:2012qa}
\hrefCMSnoop {}{A.~Belyaev, N.~D. Christensen, and A.~Pukhov, ``{CalcHEP} 3.4
  for collider physics within and beyond the standard model'',} \textit{
  Comput. Phys. Commun.} \textbf{ 184} (2013) 1729,
  \href{http://dx.doi.org/10.1016/j.cpc.2013.01.014}{\doi{10.1016/j.cpc.2013.01.014}},
\href{http://www.arXiv.org/abs/1207.6082}{\texttt{arXiv:1207.6082}}.
%%CITATION = ARXIV:1207.6082;%%.

\bibitem{Sjostrand:2014zea}
T.~Sj{\"o}strand\hrefCMSnoop {}{ {et~al.}, ``An introduction to {PYTHIA
  8.2}'',} \textit{ Comput. Phys. Commun.} \textbf{ 191} (2015) 159,
  \href{http://dx.doi.org/10.1016/j.cpc.2015.01.024}{\doi{10.1016/j.cpc.2015.01.024}},
  \href{http://www.arXiv.org/abs/1410.3012}{\texttt{arXiv:1410.3012}}.

\bibitem{Pumplin:2002vw}
J.~Pumplin\hrefCMSnoop {}{ {et~al.}, ``New generation of parton distributions
  with uncertainties from global {QCD} analysis'',} \textit{ JHEP} \textbf{ 07}
  (2002) 012,
  \href{http://dx.doi.org/10.1088/1126-6708/2002/07/012}{\doi{10.1088/1126-6708/2002/07/012}},
\href{http://www.arXiv.org/abs/hep-ph/0201195}{\texttt{arXiv:hep-ph/0201195}}.
%%CITATION = HEP-PH/0201195;%%.

\bibitem{Ball:2014uwa}
\hrefCMSnoop {}{{NNPDF} Collaboration, ``Parton distributions for the {LHC}
  {Run} {II}'',} \textit{ JHEP} \textbf{ 04} (2015) 040,
  \href{http://dx.doi.org/10.1007/JHEP04(2015)040}{\doi{10.1007/JHEP04(2015)040}},
\href{http://www.arXiv.org/abs/1410.8849}{\texttt{arXiv:1410.8849}}.
%%CITATION = ARXIV:1410.8849;%%.

\bibitem{Dreiner:2006sv}
\hrefCMSnoop {}{H.~K. Dreiner, S.~Grab, M.~Kr{\"a}mer, and M.~K. Trenkel,
  ``Supersymmetric {NLO QCD} corrections to resonant slepton production and
  signals at the {Tevatron} and the {CERN LHC}'',} \textit{ Phys. Rev. D}
  \textbf{ 75} (2007) 035003,
  \href{http://dx.doi.org/10.1103/PhysRevD.75.035003}{\doi{10.1103/PhysRevD.75.035003}},
  \href{http://www.arXiv.org/abs/hep-ph/0611195}{\texttt{arXiv:hep-ph/0611195}}.

\bibitem{Nason:2004rx}
\hrefCMSnoop {}{P.~Nason, ``A new method for combining {NLO QCD} with shower
  {Monte Carlo} algorithms'',} \textit{ JHEP} \textbf{ 11} (2004) 040,
  \href{http://dx.doi.org/10.1088/1126-6708/2004/11/040}{\doi{10.1088/1126-6708/2004/11/040}},
  \href{http://www.arXiv.org/abs/hep-ph/0409146}{\texttt{arXiv:hep-ph/0409146}}.

\bibitem{Frixione:2007vw}
\hrefCMSnoop {}{S.~Frixione, P.~Nason, and C.~Oleari, ``Matching {NLO QCD}
  computations with parton shower simulations: the {POWHEG} method'',} \textit{
  JHEP} \textbf{ 11} (2007) 070,
  \href{http://dx.doi.org/10.1088/1126-6708/2007/11/070}{\doi{10.1088/1126-6708/2007/11/070}},
  \href{http://www.arXiv.org/abs/0709.2092}{\texttt{arXiv:0709.2092}}.

\bibitem{Alioli:2010xd}
\hrefCMSnoop {}{S.~Alioli, P.~Nason, C.~Oleari, and E.~Re, ``A general
  framework for implementing {NLO} calculations in shower {Monte Carlo}
  programs: the {POWHEG BOX}'',} \textit{ JHEP} \textbf{ 06} (2010) 043,
  \href{http://dx.doi.org/10.1007/JHEP06(2010)043}{\doi{10.1007/JHEP06(2010)043}},
  \href{http://www.arXiv.org/abs/1002.2581}{\texttt{arXiv:1002.2581}}.

\bibitem{Frixione:2007nw}
\hrefCMSnoop {}{S.~Frixione, P.~Nason, and G.~Ridolfi, ``A positive-weight
  next-to-leading-order {Monte Carlo} for heavy flavour hadroproduction'',}
  \textit{ JHEP} \textbf{ 09} (2007) 126,
  \href{http://dx.doi.org/10.1088/1126-6708/2007/09/126}{\doi{10.1088/1126-6708/2007/09/126}},
  \href{http://www.arXiv.org/abs/0707.3088}{\texttt{arXiv:0707.3088}}.

\bibitem{Alwall:2014hca}
J.~Alwall\hrefCMSnoop {}{ {et~al.}, ``The automated computation of tree-level
  and next-to-leading order differential cross sections, and their matching to
  parton shower simulations'',} \textit{ JHEP} \textbf{ 07} (2014) 079,
  \href{http://dx.doi.org/10.1007/JHEP07(2014)079}{\doi{10.1007/JHEP07(2014)079}},
  \href{http://www.arXiv.org/abs/1405.0301}{\texttt{arXiv:1405.0301}}.

\bibitem{Frederix:2012ps}
\hrefCMSnoop {}{R.~Frederix and S.~Frixione, ``Merging meets matching in
  {MC@NLO}'',} \textit{ JHEP} \textbf{ 12} (2012) 061,
  \href{http://dx.doi.org/10.1007/JHEP12(2012)061}{\doi{10.1007/JHEP12(2012)061}},
  \href{http://www.arXiv.org/abs/1209.6215}{\texttt{arXiv:1209.6215}}.

\bibitem{Khachatryan:2015pea}
\hrefCMSnoop {}{{CMS Collaboration}, ``Event generator tunes obtained from
  underlying event and multiparton scattering measurements'',} \textit{ Eur.
  Phys. J. C} \textbf{ 76} (2016) 155,
  \href{http://dx.doi.org/10.1140/epjc/s10052-016-3988-x}{\doi{10.1140/epjc/s10052-016-3988-x}},
\href{http://www.arXiv.org/abs/1512.00815}{\texttt{arXiv:1512.00815}}.
%%CITATION = ARXIV:1512.00815;%%.

\bibitem{Agostinelli:2002hh}
\hrefCMSnoop {}{{GEANT4} Collaboration, ``{\GEANTfour}---a simulation
  toolkit'',} \textit{ Nucl. Instrum. Meth. A} \textbf{ 506} (2003) 250,
\href{http://dx.doi.org/10.1016/S0168-9002(03)01368-8}{\doi{10.1016/S0168-9002(03)01368-8}}.
%%CITATION = NUIMA,A506,250;%%.

\bibitem{Allison:2006ve}
\hrefCMSnoop {}{J.~Allison {et~al.}, ``{\GEANTfour} developments and
  applications'',} \textit{ IEEE Trans. Nucl. Sci.} \textbf{ 53} (2006) 270,
\href{http://dx.doi.org/10.1109/TNS.2006.869826}{\doi{10.1109/TNS.2006.869826}}.
%%CITATION = IETNA,53,270;%%.

\bibitem{ALLISON2016186}
\hrefCMSnoop {}{J.~Allison {et~al.}, ``Recent developments in {\GEANTfour}'',}
  \textit{ Nucl. Instrum. Meth. A} \textbf{ 835} (2016) 186,
\href{http://dx.doi.org/10.1016/j.nima.2016.06.125}{\doi{10.1016/j.nima.2016.06.125}}.
%%CITATION = NUIMA,A835,186;%%.

\bibitem{CMS-PRF-14-001}
\hrefCMSnoop {}{{CMS Collaboration}, ``Particle-flow reconstruction and global
  event description with the {CMS} detector'',} \textit{ JINST} \textbf{ 12}
  (2017) P10003,
  \href{http://dx.doi.org/10.1088/1748-0221/12/10/P10003}{\doi{10.1088/1748-0221/12/10/P10003}},
\href{http://www.arXiv.org/abs/1706.04965}{\texttt{arXiv:1706.04965}}.
%%CITATION = ARXIV:1706.04965;%%.

\bibitem{CMS-TDR-15-02}
\href {http://cds.cern.ch/record/2020886}{{CMS Collaboration}, ``Technical
  proposal for the {Phase-II} upgrade of the {Compact Muon Solenoid}'',} CMS
  Technical Proposal CERN-LHCC-2015-010, CMS-TDR-15-02, 2015.

\bibitem{CMS:2018jrd}
\hrefCMSnoop {}{{CMS Collaboration}, ``Performance of reconstruction and
  identification of {$\Pgt$} leptons decaying to hadrons and {$\nu_\Pgt$} in
  {$\Pp\Pp$} collisions at {$\sqrt{s}=13\TeV$}'',} \textit{ JINST} \textbf{ 13}
  (2018) P10005,
  \href{http://dx.doi.org/10.1088/1748-0221/13/10/P10005}{\doi{10.1088/1748-0221/13/10/P10005}},
  \href{http://www.arXiv.org/abs/1809.02816}{\texttt{arXiv:1809.02816}}.

\bibitem{TAU-20-001}
\hrefCMSnoop {}{{CMS Collaboration}, ``Identification of hadronic tau lepton
  decays using a deep neural network'',} 2022.
  \href{http://www.arXiv.org/abs/2201.08458}{\texttt{arXiv:2201.08458}}.
  Submitted to \textit{JINST}.

\bibitem{CMS:2014lcz}
\hrefCMSnoop {}{{CMS Collaboration}, ``{Search for physics beyond the standard
  model in dilepton mass spectra in proton-proton collisions at
  $\sqrt{s}=8\TeV$}'',} \textit{ JHEP} \textbf{ 04} (2015) 025,
  \href{http://dx.doi.org/10.1007/JHEP04(2015)025}{\doi{10.1007/JHEP04(2015)025}},
  \href{http://www.arXiv.org/abs/1412.6302}{\texttt{arXiv:1412.6302}}.

\bibitem{Pecjak:2016nee}
\hrefCMSnoop {}{B.~D. Pecjak, D.~J. Scott, X.~Wang, and L.~L. Yang, ``Resummed
  differential cross sections for top-quark pairs at the {LHC}'',} \textit{
  Phys. Rev. Lett.} \textbf{ 116} (2016) 202001,
  \href{http://dx.doi.org/10.1103/PhysRevLett.116.202001}{\doi{10.1103/PhysRevLett.116.202001}},
  \href{http://www.arXiv.org/abs/1601.07020}{\texttt{arXiv:1601.07020}}.

\bibitem{Czakon:2017wor}
M.~Czakon\hrefCMSnoop {}{ {et~al.}, ``Top-pair production at the {LHC} through
  {NNLO QCD} and {NLO EW}'',} \textit{ JHEP} \textbf{ 10} (2017) 186,
  \href{http://dx.doi.org/10.1007/JHEP10(2017)186}{\doi{10.1007/JHEP10(2017)186}},
  \href{http://www.arXiv.org/abs/1705.04105}{\texttt{arXiv:1705.04105}}.

\bibitem{CMS:2021ctt}
\hrefCMSnoop {}{{CMS Collaboration}, ``Search for resonant and nonresonant new
  phenomena in high-mass dilepton final states at {$\sqrt{s} = 13\TeV$}'',}
  \textit{ JHEP} \textbf{ 07} (2021) 208,
  \href{http://dx.doi.org/10.1007/JHEP07(2021)208}{\doi{10.1007/JHEP07(2021)208}},
  \href{http://www.arXiv.org/abs/2103.02708}{\texttt{arXiv:2103.02708}}.

\bibitem{CMS:2019ied}
\hrefCMSnoop {}{{CMS Collaboration}, ``{Performance of the reconstruction and
  identification of high-momentum muons in proton-proton collisions at
  $\sqrt{s} = 13\TeV$}'',} \textit{ JINST} \textbf{ 15} (2020) P02027,
  \href{http://dx.doi.org/10.1088/1748-0221/15/02/P02027}{\doi{10.1088/1748-0221/15/02/P02027}},
  \href{http://www.arXiv.org/abs/1912.03516}{\texttt{arXiv:1912.03516}}.

\bibitem{Butterworth:2015oua}
\hrefCMSnoop {}{J.~Butterworth {et~al.}, ``{PDF4LHC} recommendations for {LHC
  Run II}'',} \textit{ J. Phys. G} \textbf{ 43} (2016) 023001,
  \href{http://dx.doi.org/10.1088/0954-3899/43/2/023001}{\doi{10.1088/0954-3899/43/2/023001}},
  \href{http://www.arXiv.org/abs/1510.03865}{\texttt{arXiv:1510.03865}}.

\bibitem{Khachatryan:2016kdb}
\hrefCMSnoop {}{{CMS Collaboration}, ``Jet energy scale and resolution in the
  {CMS} experiment in {$\Pp\Pp$} collisions at {$8\TeV$}'',} \textit{ JINST}
  \textbf{ 12} (2017) P02014,
  \href{http://dx.doi.org/10.1088/1748-0221/12/02/P02014}{\doi{10.1088/1748-0221/12/02/P02014}},
\href{http://www.arXiv.org/abs/1607.03663}{\texttt{arXiv:1607.03663}}.
%%CITATION = ARXIV:1607.03663;%%.

\bibitem{CMS:2021xjt}
\hrefCMSnoop {}{{CMS Collaboration}, ``Precision luminosity measurement in
  proton-proton collisions at {$\sqrt{s} = 13\TeV$} in 2015 and 2016 at
  {CMS}'',} \textit{ Eur. Phys. J. C} \textbf{ 81} (2021) 800,
  \href{http://dx.doi.org/10.1140/epjc/s10052-021-09538-2}{\doi{10.1140/epjc/s10052-021-09538-2}},
  \href{http://www.arXiv.org/abs/2104.01927}{\texttt{arXiv:2104.01927}}.

\bibitem{CMS:2018elu}
\href {https://cds.cern.ch/record/2621960}{{CMS Collaboration}, ``{CMS}
  luminosity measurement for the 2017 data-taking period at {$\sqrt{s} =
  13\TeV$}'',} CMS Physics Analysis Summary CMS-PAS-LUM-17-004, 2018.

\bibitem{CMS:2019jhq}
\href {https://cdsweb.cern.ch/record/2676164}{{CMS Collaboration}, ``{CMS}
  luminosity measurement for the 2018 data-taking period at {$\sqrt{s} =
  13\TeV$}'',} CMS Physics Analysis Summary CMS-PAS-LUM-18-002, 2019.

\bibitem{Barlow:1993dm}
\hrefCMSnoop {}{R.~J. Barlow and C.~Beeston, ``Fitting using finite {Monte
  Carlo} samples'',} \textit{ Comput. Phys. Commun.} \textbf{ 77} (1993) 219,
  \href{http://dx.doi.org/10.1016/0010-4655(93)90005-W}{\doi{10.1016/0010-4655(93)90005-W}}.

\bibitem{CowanPDGStat}
\hrefCMSnoop {}{{G. Cowan, ``Statistics'', Ch. 39 in Particle Data Group},
  C.~Patrignani {et~al.}, ``Review of particle physics'',} \textit{ Chin. Phys.
  C} \textbf{ 40} (2016) 100001,
\href{http://dx.doi.org/10.1088/1674-1137/40/10/100001}{\doi{10.1088/1674-1137/40/10/100001}}.
%%CITATION = CHPHD,C40,100001;%%.

\bibitem{ATLAS:2011tau}
\href {https://cds.cern.ch/record/1379837}{{ATLAS and CMS Collaborations, and
  the LHC Higgs Combination Group}, ``Procedure for the {LHC Higgs} boson
  search combination in summer 2011'',} Technical Report CMS-NOTE-2011-005,
  ATL-PHYS-PUB-2011-11, 2011.

\bibitem{Baak:2014fta}
\hrefCMSnoop {}{M.~Baak, S.~Gadatsch, R.~Harrington, and W.~Verkerke,
  ``Interpolation between multi-dimensional histograms using a new non-linear
  moment morphing method'',} \textit{ Nucl. Instrum. Meth. A} \textbf{ 771}
  (2015) 39,
  \href{http://dx.doi.org/10.1016/j.nima.2014.10.033}{\doi{10.1016/j.nima.2014.10.033}},
  \href{http://www.arXiv.org/abs/1410.7388}{\texttt{arXiv:1410.7388}}.

\bibitem{Allanach:2007qj}
\hrefCMSnoop {}{B.~C. Allanach and C.~G. Lester, ``Sampling using a `bank' of
  clues'',} \textit{ Comput. Phys. Commun.} \textbf{ 179} (2008) 256,
  \href{http://dx.doi.org/10.1016/j.cpc.2008.02.020}{\doi{10.1016/j.cpc.2008.02.020}},
  \href{http://www.arXiv.org/abs/0705.0486}{\texttt{arXiv:0705.0486}}.

\end{thebibliography}\endgroup

\cleardoublepage \appendix\section{The CMS Collaboration \label{app:collab}}\begin{sloppypar}\hyphenpenalty=5000\widowpenalty=500\clubpenalty=5000
\cmsinstitute{Yerevan Physics Institute, Yerevan, Armenia}
{\tolerance=6000
A.~Tumasyan\cmsAuthorMark{1}\cmsorcid{0009-0000-0684-6742}
\par}
\cmsinstitute{Institut f\"{u}r Hochenergiephysik, Vienna, Austria}
{\tolerance=6000
W.~Adam\cmsorcid{0000-0001-9099-4341}, J.W.~Andrejkovic, T.~Bergauer\cmsorcid{0000-0002-5786-0293}, S.~Chatterjee\cmsorcid{0000-0003-2660-0349}, K.~Damanakis\cmsorcid{0000-0001-5389-2872}, M.~Dragicevic\cmsorcid{0000-0003-1967-6783}, A.~Escalante~Del~Valle\cmsorcid{0000-0002-9702-6359}, R.~Fr\"{u}hwirth\cmsAuthorMark{2}\cmsorcid{0000-0002-0054-3369}, M.~Jeitler\cmsAuthorMark{2}\cmsorcid{0000-0002-5141-9560}, N.~Krammer\cmsorcid{0000-0002-0548-0985}, L.~Lechner\cmsorcid{0000-0002-3065-1141}, D.~Liko\cmsorcid{0000-0002-3380-473X}, I.~Mikulec\cmsorcid{0000-0003-0385-2746}, P.~Paulitsch, F.M.~Pitters, J.~Schieck\cmsAuthorMark{2}\cmsorcid{0000-0002-1058-8093}, R.~Sch\"{o}fbeck\cmsorcid{0000-0002-2332-8784}, D.~Schwarz\cmsorcid{0000-0002-3821-7331}, S.~Templ\cmsorcid{0000-0003-3137-5692}, W.~Waltenberger\cmsorcid{0000-0002-6215-7228}, C.-E.~Wulz\cmsAuthorMark{2}\cmsorcid{0000-0001-9226-5812}
% ...
\par}
\vskip\cmsinstskip
\dag:~Deceased\\
$^{1}$Also at Yerevan State University, Yerevan, Armenia\\
$^{2}$Also at TU Wien, Vienna, Austria\\
% ...
\end{sloppypar}
%%% END EDITABLE REGION %%%
% skeleton_end
\end{document}